\documentclass{emulateapj}

\usepackage{graphicx}
\usepackage{datetime}

\slugcomment{To appear in ApJ.}

\shorttitle{Dead Zone or Planet Carved Gap Edges?}
\shortauthors{Reg\'aly et al.}

\usepackage{color}

\begin{document}

\newcommand{\ajc}[1]{{\bf #1} }

\title{Interpreting Brightness Asymmetries in Transition Disks:\\ 
Vortex at Dead Zone or Planet-carved Gap Edges?}

\author{Zs. Reg\'aly$^{1}$\thanks{E-mail:regaly@konkoly.hu}, A. Juh\'asz$^{2,3}$, D. Neh\'ez$^{1,4}$}
\affil{$^1$Konkoly Observatory, Research Center for Astronomy and Earth Sciences, Hungarian Academy of Sciences,
\\1121, Budapest, Konkoly Thege Mikl\'os \'ut 15-17, Hungary, e-mail:regaly@konkoly.hu}
\affil{$^2$Leiden Observatory, Leiden University, P.O. Box 9513, NL-2300 RA Leiden, The Netherlands}
\affil{$^3$Institute of Astronomy, University of Cambridge, Madingley Road, Cambridge CB3 0HA, UK}
\affil{$^4$Department of Astronomy, E\"otv\"os Lor\'and University, Budapest, Hungary, H-1117}

 
\begin{abstract}
Recent submillimeter observations show nonaxisymmetric brightness distributions with a horseshoe-like morphology for more than a dozen transition disks. The most-accepted explanation for the observed asymmetries is the accumulation of dust in  large-scale vortices. Protoplanetary disks vortices can form by the excitation of Rossby wave instability in the vicinity of  a steep pressure gradient, which can develop at the edges of a giant planet-carved gap or at the edges of an accretionally inactive zone. We studied the formation and evolution of vortices formed in these two distinct scenarios by means of two-dimensional locally isothermal hydrodynamic simulations. We found that the vortex formed at the edge of a planetary gap is short-lived, unless the disk is nearly inviscid. In contrast, the vortex formed at the outer edge of a dead zone is long-lived. The vortex morphology can be significantly different in the two scenarios: the vortex radial and azimuthal extensions are $\sim1.5$ and $\sim3.5$ times larger for the dead-zone edge compared to gap models. In some particular cases, the vortex aspect ratios can be similar in the two scenarios; however, the vortex azimuthal extensions can be used to distinguish the vortex formation mechanisms. We calculated predictions for vortex observability in the submillimeter continuum with ALMA. We found that the azimuthal and radial extent of the brightness asymmetry correlates with the vortex formation process within the limitations of $\alpha$-viscosity prescription.
\end{abstract}

\keywords{accretion, accretion disks --- protoplanetary disks --- hydrodynamics --- methods: numerical}

\section{Introduction}

Transition disks are believed to be the bridging phase between gas-rich primordial disks and gas-poor debris disks. The characteristic feature of transition disks is the deficit of infrared excess in their spectral energy distribution \citep{Stormetal1989,Skrutskieetal1990}, which is attributed to the formation of dust-depleted gaps, cavities. Such cavities depleted of dust by 5-6 orders of magnitude have already been confirmed by submillimeter imaging (see a review in \citealp{Espaillatetal2014}).

Recent spatially resolved submillimeter observations of transition disks revealed nonaxisymmetric brightness distributions for more than a dozen cases (see, e.g., \citealt{Andrewsetal2009,Andrewsetal2011,Brownetal2009, Hughesetal2009,Isellaetal2010,Mathewsetal2012, Tangetal2012,Casassusetal2013,Casassusetal2015,Fukagawaetal2013,vanderMareletal2013,Perezetal2014,Hashimotoetal2015,Marinoetal2015,Momoseetal2015,Wrightetal2015}). The observed lopsided brightness distributions have in most cases a horseshoe-like shape. Since the disk is believed to be optically thin at submillimeter wavelengths, these horseshoe-shaped asymmetries reflect perturbation in the density and/or temperature structure of the disk. The observed asymmetries show remarkable agreement with the morphology of large-scale anticyclonic vortices predicted to form in protoplanetary disks  \citep{Regalyetal2012}. The accumulation of millimeter- to centimeter-sized dust has now becOme the commonly accepted explanation for the observed asymmetries.

Formation of a large-scale vortex can be triggered in protoplanetary disks due to baroclinic instability \citep{KlahrBodenheimer2003,LyraKlahr2011,Raettigetal2013,Lyra2014} or to the Rossby wave instability (RWI) via the coagulation of smaller-scale vortices \citep{Lovelaceetal1999, Lietal2000, Lietal2001, Lyraetal2009b,Meheutetal2010, Meheutetal2012a, Meheutetal2012b, Meheutetal2012c, Meheutetal2013, Richardetal2013,Flocketal2015}.

\citet{BargeSommeria1995} and \citet{KlahrHenning1997} have already shown that particles tend to get trapped in anticyclonic vortices. Particles of about a centimeter to a meter in size drifting into the pressure maxima, i.e., to the core of the vortices \citep{Lyraetal2009b,Dzyurkevichetal2010,Katoetal2010} may form gravitationally bound clumps of solids, which can coalesce forming embryos between the masses of the Moon and Mars. The swarm of these embryos can evolve further by mutual collisions forming massive cores ($\sim10\,M_\oplus$) of giant planets in $\sim5\times10^5$\,yr \citep{Sandoretal2011}. 

The radial pressure gradient at the  location of the vortex can halt not only dust grains but also larger bodies migrating in the type I regime. However, trapping of low-mass protoplanets (about $10M_\oplus$) is  found to be only temporary \citep{Regalyetal2013}. Large-scale vortices are known to accelerate the formation of planetesimals and planetary embryos in the core accretion paradigm; thus, such vortices are extremely important for planet formation, helping to overcome the meter-size barrier problem \citep{BlumWurm2008}, and the time-scale problem of oligarchic growth \citep{Thommesetal2003}. Moreover, \citet{KlahrBodenheimer2006} have also shown that vortices can hasten the formation of giant planets by decreasing the time to form massive cores ($\sim10-15M_\oplus$) required for launching runaway gas accretion. From this perspective, these vortices can be regarded as "planetary cradles."

While the presence of such vortices in disks seems to be widely accepted, less is known about their formation and lifetime. In this paper, we study the formation and evolution of vortices excited by RWI. The RWI can be excited in protoplanetary disks in various situations, e.g., due to a steep pressure gradient. Such pressure gradients can form, e.g. at the edges of a gap opened by an embedded planet  \citep{Lietal2005}, at the edges of the disk's accretionally inactive dead zone \citep{Lovelaceetal1999,Lietal2000}, or in protostellar infall from the natal cloud near the centrifugal radius \citep{Baeetal2015}. However, so far, vortices have only been seen in class II disks, which no longer have any envelope, so this scenario surely does not apply to them.

Both the large-scale vortex formation scenarios have difficulties explaining the asymmetric brightness features of transition disks. Formation of a large-scale vortex at the edge of a giant planet-carved gap is found to be delayed or even suppressed by the disk self-gravity for a sufficiently high-mass ($M_\mathrm{disk}/M_*\gtrsim0.38$) disk \citep{LinPapaloizou2011}. In shearing sheet simulations, vortices are found to be transient structures due to the effect of self-gravity, inhibiting the formation of a single large-scale vortex \citep{MamatsashviliRice2009}. If RWI is excited, vortices may also have short lifetimes due to the disk viscosity \citep{deValBorroetal2007,Ataieeetal2013,Fuetal2014b}. Moreover, vortices can be destroyed due to the effect of dust feedback if the local dust-to-gas mass ratio approaches unity \citep{Johansenetal2004,InabaBarge2006,Lyraetal2009a,Fuetal2014a}. However, \citet{Raettigetal2015} showed that the streaming instability that causes vortex decay is only a temporary effect, and vortices can be re-established. Excitation of vortices at the dead-zone edge by RWI seems to require a sharper viscosity transition \citep{Lyraetal2009b,Regalyetal2012} than it is expected to form at the outer dead-zone edge \citep{Dzyurkevichetal2013}. Note, however, that  \citet{Lyraetal2015} found that a smooth change in gas resistivity does not imply a smooth transition in turbulent stress, in which case a large-scale vortex can form, although the resistivity transition is  smooth.  Considering the abovementioned difficulties, to understand the origin of large-scale brightness asymmetries of transition disks, we must rely on observations and compare them to the predictions given by vortex formation models.

In this paper, we investigate whether or not we can infer the formation mechanism of the vortex from the morphology of the disk as seen in submillimeter ALMA observations. In Section\,2, we model long-term evolution of RWI-excited large-scale vortices by means of locally isothermal 2D hydrodynamical simulations for a wide range of disk parameters and a morphological comparison of vortex structures in the gas distribution. In Section\,3, we present submillimeter images calculated for ALMA observations based on hydrodynamical results. In Section\,4, we provide a discussion on observed versus synthetic submillimeter brightness asymmetries and a review of our model caveats. The paper closes (Section\,5) with a summary of our findings and concluding remarks.

\section{Formation of Large-scale Vortex}

\begin{figure*}
	\centering
	\includegraphics[width=15cm]{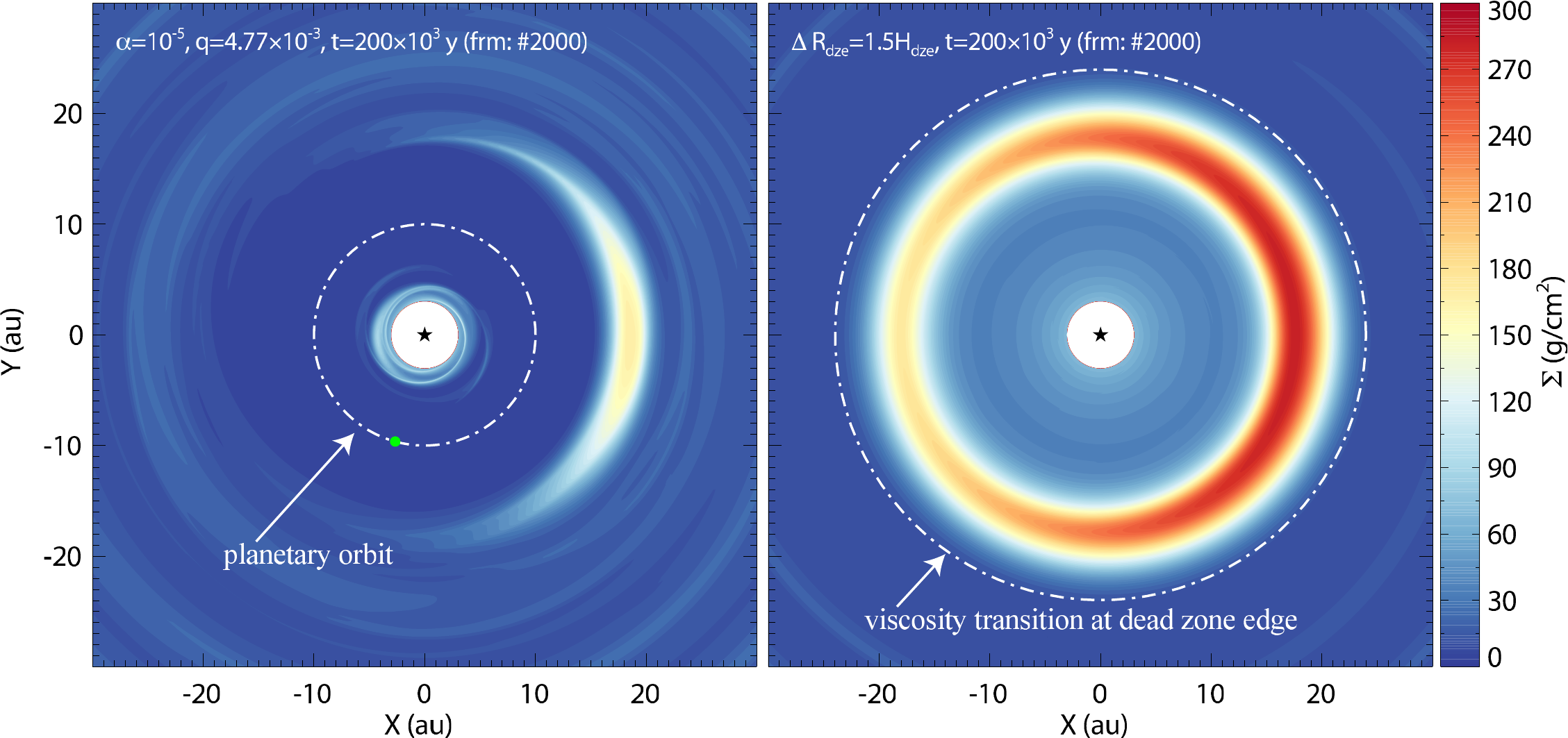}
	\caption{Density snapshots of hydrodynamic simulations taken at the end of the simulations ($t=200\times10^3$\,yr corresponding to $\sim2600$ vortex orbits). Left hand panel: GAP model in which a gap is carved by a $5M_\mathrm{Jup}$ giant planet orbiting at 10\,au. Right hand panel: DZE model assuming $\alpha=10^{-2}$, $\alpha_\mathrm{dz}=10^{-4}$, $R_\mathrm{dze}=24$\,au, and $\Delta R_\mathrm{dze}=1.5H_\mathrm{dze}$. }
	\label{fig:vortex-evol-2D}
\end{figure*}

\subsection{Hydrodynamic Models}

We investigated the formation and evolution of a large-scale vortex in two scenarios. In the first scenario, the vortex forms at the edges of a gap opened by a giant planet. Note that we only model the vortex formation at the outer gap edge.\footnote{Since the vortex evolution at the inner gap edge can be strongly affected by the simulation domain's inner boundary, we do not follow the evolution of vortices at the inner gap edge.} In the second scenario, the vortex is excited by a steep transition in the viscosity, e.g. at the edge of a dead zone. Two-dimensional grid-based, global hydrodynamic disk simulations have been used to study the properties of the vortices under these conditions. We used the GPU-supported version of the code {\small FARGO} \citep{Masset2000}, which numerically solves the vertically integrated continuity and Navier--Stokes equations, using locally isothermal approximation in a cylindrical coordinate system centered on the star.  

We used the $\alpha$ prescription for the viscosity  \citep{ShakuraSunyaev1973} with $10^{-5}\leq\alpha\leq10^{-3}$. Initially, the gas density has a power-law profile $\Sigma(R)=\Sigma_0R^{-p}$, with $p=0.5,\,1$ and $1.5$ and $\Sigma_0=1.45\times10^{-5},\,5.3\times10^{-4}$, and $1.45\times10^{-4}$, respectively, for which cases the disk mass is $0.02\,M_*$ independent of $p$.  As the Toomre $Q$ parameter \citep{Toomre1964} initially is higher than unity ($Q_\mathrm{min}\simeq13$ at the vortex radial distance) for the investigated density profiles or even in the center of the fully developed large-scale vortex ($Q_\mathrm{min}\simeq5$), the disk is gravitationally stable; therefore, we neglect the disk self-gravity.

The pressure scale height of the disk is assumed to have a power-law dependence on the  radius, $H(R) = hR^{1+\gamma}$, where $h$ is the aspect ratio and $\gamma$ is the flaring index.  For flat-disk assumption, we use $h=0.05$ and $\gamma=0$, and for flaring-disk models, we use $h=0.05$ and $\gamma=2/7$ \citep{ChiangGoldreich1997}. We define a fiducial disk setup for which case $p=0.5$, $\gamma=0$, and $h=0.05$ are used.

Assuming locally isothermal approximation, the disk's temperature profile is $T(r)\sim R^{2(\gamma-0.5)}$. In this approximation, the equation of state of gas is $P=c_\mathrm{s}(R)\Sigma(R)$, where $c_\mathrm{s}(R)=\Omega(R) H(R)$ is the sound speed $\Omega(R)$ and $H(R)$ are the local Keplerian angular velocity and the local pressure scale height, respectively. The unit length is taken to be 1\,au and the unit mass is the stellar mass. Assuming that the unit time is the inverse of the Kepler frequency, the orbital period becomes 2$\pi$ and the gravitational constant is unity. 

The spatial extension of the computational domain is $3\textrm{--}50$\,au, consisting of $N_R = 512$ logarithmically distributed radial and $N_\phi=1024$ equidistant azimuthal grid cells. At the inner and outer boundaries, a wave-damping boundary condition is applied \citep{deValBorroetal2006}. For this boundary condition, the disk mass is not conserved. Indeed, the disk mass increases with time in the simulation, but the total variation of the disk mass during the entire simulation is less than a percent. Note that setting the boundary conditions to open has no significant effect on the results but increase the computational time due to waves developed near the inner boundary. In planet-bearing models, we use $\epsilon H(a)$ as the smoothing of the gravitational potential of the planet with an appropriate value of $\epsilon=0.6$ \citep{Kleyetal2012}.

In our simulations, the star is always at the center of the numerical domain. Therefore, the indirect potential is taken into account (see its importance in \citealp{MittalChiang2015,ZhuBaruteau2016}); i.e., two additional indirect potentials are added to the total potential due to (1) the shift of the barycenter by the disk material (i.e. large density asymmetry caused by the vortex) and (2) the shift of the barycenter due to the massive planet. 

Simulations cover $200\times10^{3}$\,yr, which corresponds to $\sim6350$ orbits of the nonmigrating  planet. Since the initial parameters are set such that the vortex forms at $\sim18$~au, the simulation time corresponds to $\sim2600$ Keplerian orbits at the distance of the vortex center.

\subsection{Vortex Development at the Gap Outer Edge}
\label{sect:GAP-model}

To study the effect of the planet mass on the formation and evolution of vortices developed at the gap edge, we set the planet-to-star mass ratio to $q=1.19,\,2.38,\,4.77,$ and $9.54\times10^{-3}$, corresponding to a planet mass of $1.25,\,2.5,\,5,$ and $10\,M_\mathrm{Jup}$ assuming a solar-mass central star. We assume  constant semi major axis of $a_\mathrm{p}=10$\,au for the planet.  We also run simulations in three different viscosity regimes for $q=4.77\times10^{-3}$ giant planet, where $\alpha=10^{-3},\,10^{-4}$, and $10^{-5}$ are assumed. In the following, we refer to these models as "GAP models".

A gap is opened by the planet in all simulations because the gap-opening criterion is satisfied as long as $\alpha\lesssim0.01$ for the assumed disk aspect ratio and planet-to-star mass ratio \citep{Cridaetal2006}. The RWI is excited in all models in about 100 orbits, initially with mode number $m=5$ being the fastest growth mode, as predicted by analytic theory for relatively low viscosity ($\alpha\leq10^{-3}$;  \citealp{Lovelaceetal1999}).

As the vortices are subject to a coalescing process, a large-scale vortex develops in a couple of additional planetary orbits, which is due to the domination of lower mode numbers (large vortices)  after the saturation of the initially linear growth of the $m=5$ mode \citep{GodonLivio1999}. However, the vortex decays due to the viscous evolution of the disk (see, e.g., \citealp{Cridaetal2006}). For $\alpha=10^{-3}$ and $10^{-4}$, the lifetime of the vortex,  $\tau_\mathrm{v}$, is shorter than $11\times10^{3}$\,yr and $110\times10^3$\,yr, corresponding to about 350 and 3500 planetary (or $144$ and $1440$ vortex) orbits, respectively. 

We run several models to explore the effect of the disk parameters $p$, $\gamma$ or planetary mass on $\tau_\mathrm{v}$ in the $\alpha=10^{-4}$ model. The value of $\tau_\mathrm{v}$ decreases for steeper surface density distributions as it is found to be $\sim90\times10^{3}$\,yr and $\sim60\times 10^3$\,yr (corresponding to $\sim1100$ and $\sim785$ vortex orbits) for $p=1.0$ and $1.5$, respectively. It is known that the indirect potential (emerged due to the barycenter shift) has the side effect of increasing the vortex strength and therefore its lifetime \citep{MittalChiang2015,ZhuBaruteau2016,RegalyVorobyov2017}. Since the mass confined inside the vortex decreases with increasing $p$, if the disk mass is kept constant, the indirect potential being proportional to the mass of the asymmetry is also decreases.

For the flaring-disk model, where $\gamma=2/7$ and $p=0.5$, $\tau_\mathrm{v}\simeq6.5\times10^3$\,yr (corresponding to about 200 planetary and $\sim85$ vortex orbits). This can be explained by that the viscosity $\nu_\mathrm{flared}\simeq5\nu_\mathrm{flat}$ at $R\simeq18$\,au, which results in a weaker vortex and faster viscous evolution of the gap edge.

Assuming the $p=0.5$ flat-disk model,  $\tau_\mathrm{v}$ is found to be correlated with the planetary mass, i.e. $\tau_\mathrm{v}\simeq30,\,55,\,110$, and $150\times10^3$\,yr  (corresponding to $\sim393,\,720,\,1440$, and $1964$ vortex orbits) for a $1.25,\,2.5,\,5$, and $10\,M_\mathrm{Jup}$ planet, respectively. This can be explained by the fact that the density gradient, and therefore the pressure gradient, is sharper for a higher planet mass, which results in a stronger vortex and longer vortex decay time.

Contrary to the previous cases, the vortex lifetime is longer than the simulations for $\alpha=10^{-5}$ models with $p=0.5$ and $1$, while the vortex is dissipated by the end of the simulation for $p=1.5$. The left panel of Fig.\,\ref{fig:vortex-evol-2D} shows the density distribution of the disk in the GAP model at the end of the simulation for $\alpha=10^{-5}$ assuming $p=0.5$. We emphasise that the vortex can survive for a sufficiently long time to be observed in the submillimeter (let us say $\gg100\times10^{3}$\,yr by the time the transitional disk phase presumably develops) only for a nearly inviscid assumption ($\alpha\leq10^{-5}$) in conjunction with the results of previous investigations \citep{GodonLivio1999,deValBorroetal2007, Ataieeetal2013,Fuetal2014a}.

\subsection{Vortex Evolution at the Dead-zone Outer Edge}

To model the formation of vortices in a disk with a dead zone, we reduced the value of $\alpha$ smoothly within a certain radius  such that $\alpha(R)=\alpha\delta_\alpha(R)$. The viscosity reduction factor is given by 
\begin{equation}
	\label{eq:deltaalpha}
	\delta_\alpha(R)=1-\frac{1}{2}\left(1-\alpha_\mathrm{mod}\right)\left[1-\tanh\left(\frac{R-R_\mathrm{dze}}{\Delta R_\mathrm{dze}}\right)\right],
\end{equation}
where $\alpha_\mathrm{mod}=0.01$ is the depth of turbulent viscosity reduction. To quantify the dead-zone edge radius, $R_\mathrm{dze}=24$\,au was used, where we adopt the results of \citet{MatsumuraPudritz2005}, who found that $R_\mathrm{dze}$ lies between 12 and 36\,au, depending on the density of the disk. For the dead-zone edge width, we assumed $\Delta R_\mathrm{dze}=1$ and $1.5H_\mathrm{dze}$, where $H_\mathrm{dze}=R_\mathrm{dze}h$ (assuming a flat-disk model) is the disk scale-height at the outer edge of the dead zone. Although these parameters correspond to 1.8 and 2.7\,au assuming $h=0.05$, according to the Equation (\ref{eq:deltaalpha}), the gross widths of the regions, where the viscosity transitions occur are 3.6 and 5.4\,au. Note that we did not study the effect of the inner edge of the dead zone. In the following, we refer to these models as "DZE models".

Due to the sudden change in viscosity (i.e. in the accretion rate) at the outer dead-zone edge, density and pressure enhancements appear that are found to be unstable to RWI \citep{VarniereTagger2006, Terquem2008}. The fastest-growing mode is $m=5$; thus, five anticyclonic vortices initially form that later coalesce \citep{Lovelaceetal1999,Lietal2000}. As a result, a large-scale vortex develops inside the dead zone. Note that the large-scale vortex forms at about the same distance, $R\simeq18$\,au, as in the GAP models assuming $R_\mathrm{dze}=24$\,au (see Fig.\,\ref{fig:vortex-evol-2D}). DZE vortices are also subject to slow decay due to the disk viscous evolution similarly to the GAP models. However, the lifetime of the large-scale vortices are found to be longer than the simulation time in all models; i.e., DZE vortices are long-lived structures for these cases \citep{Regalyetal2012,Lin2014}.

\subsection{Comparison of  GAP and DZE Vortices}
\label{sect:hydro-morph}

 \begin{figure}
	\centering
	\includegraphics[width=\columnwidth]{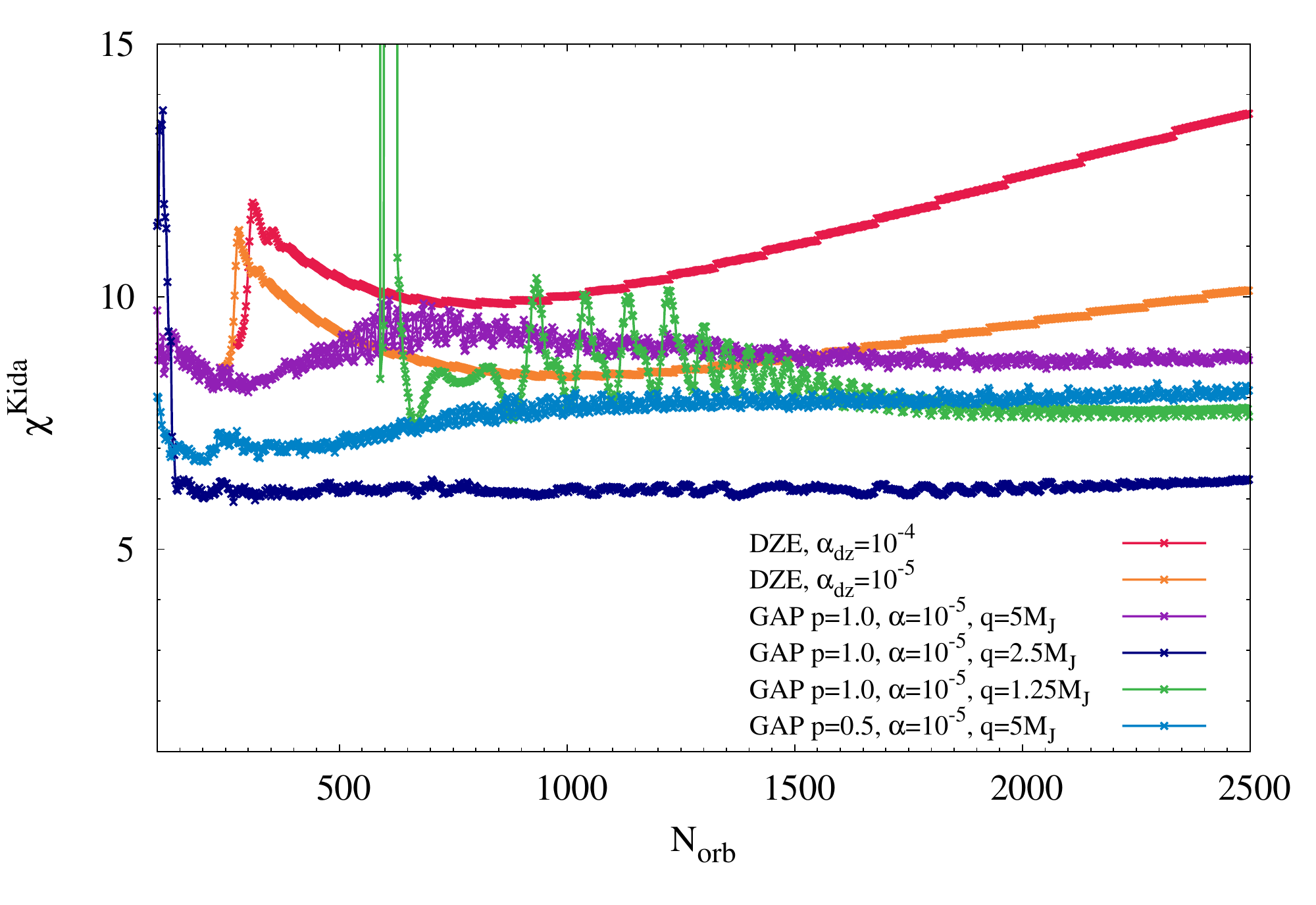}
	\caption{Evolution of  the vortex aspect ratio assuming the Kida model (Equation\,\ref{eq:Kida}) as a function of the number of the vortex orbits in six different DZE and GAP models. Since the outer spiral wave excited by the planet enters to the vortex periodically, $\chi$ is smoothed by a running average for GAP models.}
	\label{fig:chi}
\end{figure}

\begin{figure*}
	\centering
	\includegraphics[width=8.8cm]{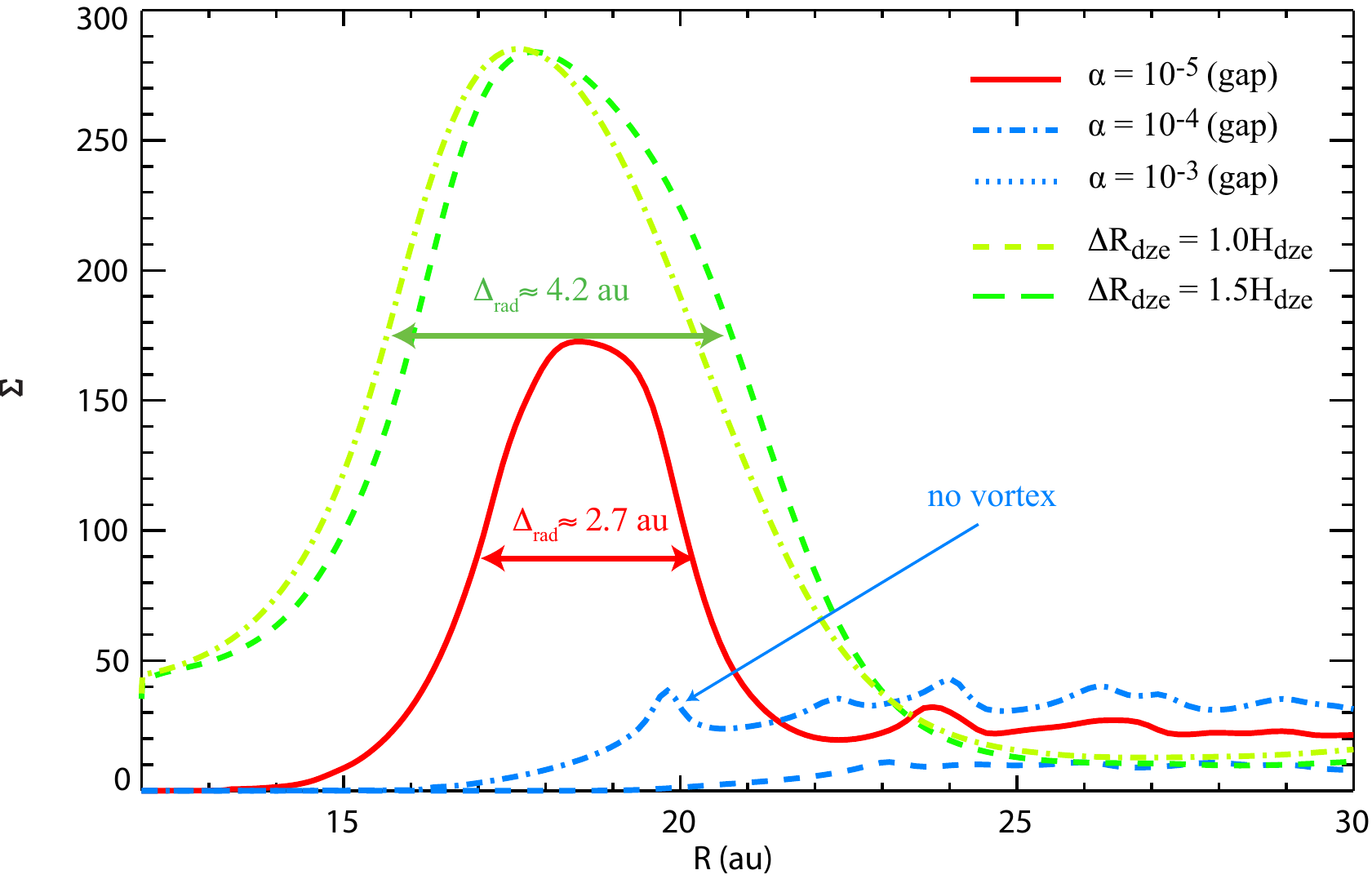}
	\includegraphics[width=8.6cm]{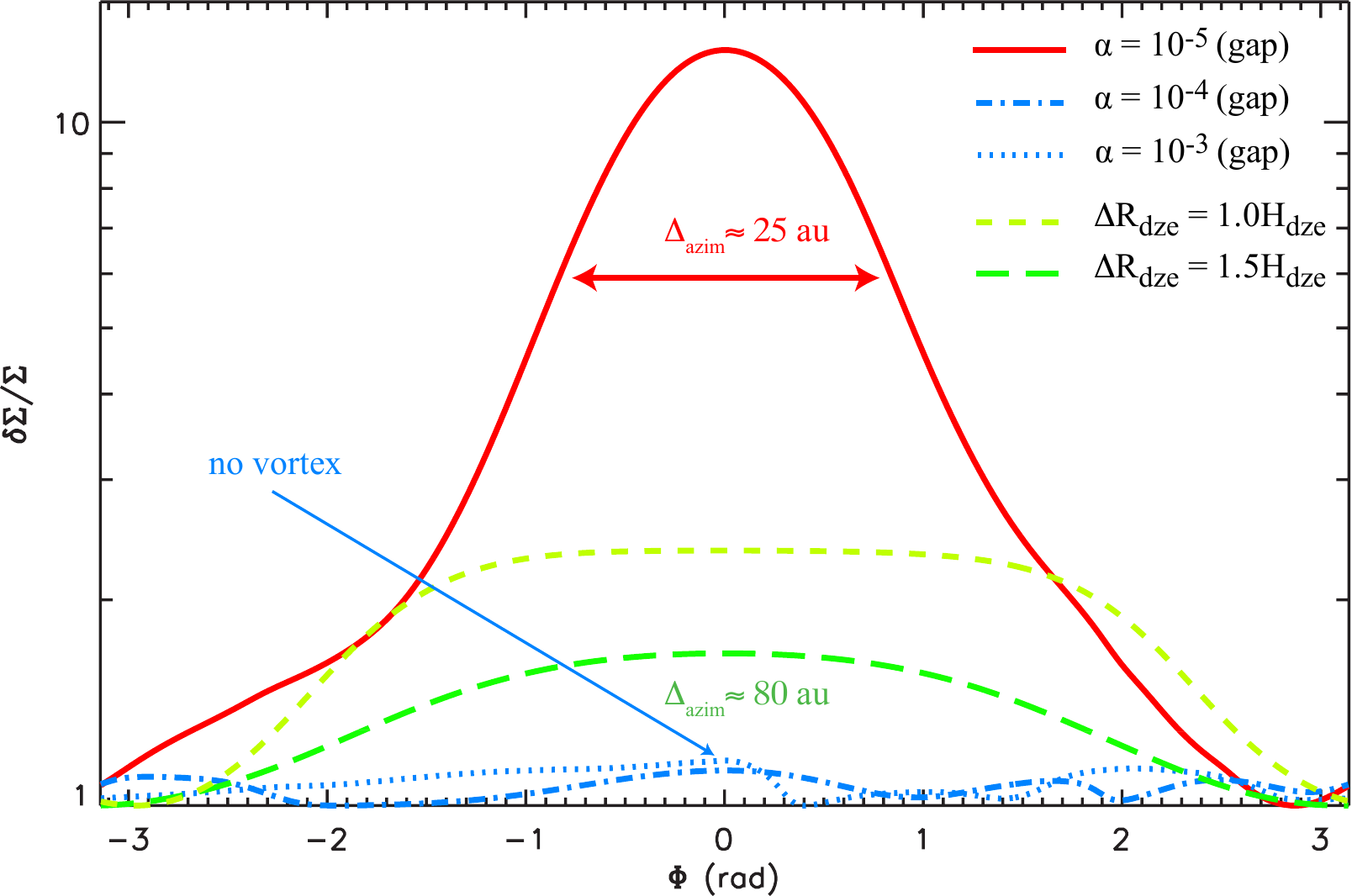}
	\caption{Radial (left panel) and azimuthal (right panel) density profiles across the vortex center taken at the end of the simulations in the GAP and DZE models. The vortex widths are determined at the $1/e$ level of the density maxima at each profile.}
	\label{fig:dens-profiles}
\end{figure*}

To compare the morphologies of the full-fledged vortices in the GAP and DZE models, we determine the vortex strengths characterized by the Rossby number,
\begin{equation}
	R_\mathrm{o}(R,\phi)=\frac{\nabla\times({\bf u}(R,\phi)-R\Omega{(R)})}{2\Omega(R_\mathrm{v})},
	\label{eq:Rossby-number}
\end{equation}
i.e. the $z$ component of the vorticity measured in the local frame of the vortex divided by the global vorticity of the Keplerian disk.  The vortex strength, $R_\mathrm{o}$, defined by the mean value of $R_\mathrm{o}(r,\phi)$ inside the vortex (set by the contour level of density at $1/e$ times the maximum density measured at the vortex center), is found to be $-0.14$ and $-0.13$ for $p=0.5$ and $1$ in the GAP models and $-0.068$, $-0.076$, and $-0.041$ for $p=0.5$, $1$, and $1.5$ in the DZE models at the end of the simulations.

For a steady-state vortex with uniform vorticity,
\begin{equation}
	R_o^\mathrm{Kida}=-\frac{3}{4}\frac{\chi^2+1}{\chi(\chi-1)}+\frac{3}{4},
	\label{eq:Kida}
\end{equation}
where $\chi$ is the vortex aspect ratio defined by the ratio of the azimuthal and radial axes of the vortex \citep{Kida1981,Chavanis2000}. For a vorticity field whose strength linearly depends on the distance from the vortex center, \citet{Goodmanetal1987} proposed 
\begin{equation}
	R_o^\mathrm{GNG}=-\frac{\sqrt{3}}{2}\frac{\chi^2+1}{\chi\sqrt{\chi^2-1}}+\frac{3}{4},
	\label{eq:GNG}
\end{equation}
while \citet{SurvilleBarge2015} recently proposed a more elaborate Gaussian model, where
\begin{equation}
	R_o^\mathrm{Gauss}=-\frac{1}{2}\frac{\chi^2+1}{\chi^2-1}\left(\frac{3}{2}-\sqrt{3}\right)+\frac{3}{2}.
	\label{eq:Gauss}
\end{equation}
We emphasize that Equations~(\ref{eq:GNG}) and (\ref{eq:Gauss}) give $R_o=(1/4)(3-2\sqrt{3})<-0.116$, when $\chi\rightarrow\infty$. To compare vortex aspect ratios for the GAP and DZE vortices, we use the Kida approximation, as $R_\mathrm{o}$ is found to be $\geq-0.1$ in all DZE models. 

The evolution of  $\chi$ calculated according to Equations~(\ref{eq:Rossby-number}) and (\ref{eq:Kida}) is shown in Fig.~\ref{fig:chi} for the DZE and GAP models.  In the $\alpha=10^{-5}$ GAP models, $\chi$ is found to be $6\lesssim\chi\lesssim8$ and shows only weak evolution in time. However, the vortex lifetime is shorter than the simulation in the $\alpha=10^{-4}$ GAP models, and $\chi\gtrsim10$. Note that $\chi$ shows no clear trends with the planet mass, as we found that the strongest vortex, $\chi\simeq6$, formed in models where a $q=2.5\,M_\mathrm{Jup}$ planet is assumed (dark blue curve in Fig.~\ref{fig:chi}). In contrast, $\chi$ evolves strongly in time for the DZE models: initially, the vortex strengthens, i.e. $\chi$ decreases up to $\sim500$ orbits; then it starts to weaken, i.e., $\chi$ increases, $\chi_\mathrm{max}\simeq14$ and $10$ for $\alpha_\mathrm{dz}=10^{-4}$ and $10^{-5}$, respectively. Note that $\chi$ does not saturate by the end of the simulation for the DZE models. However, as we showed earlier, a quasi-stationary vortex mode can be reached after $\sim5000$ orbits of the vortex (see Figure\,5 in \citealp{Regalyetal2012}). In summary, $\chi\simeq8$ in the GAP models where the vortex is maintained until the end of the simulations assuming $\alpha\leq10^{-5}$. For the DZE models, the vortex weakens by the end of the simulation and thus has a larger aspect ratio: $\chi\simeq14$ and $\chi\simeq10$ for $\alpha_\mathrm{dz}=10^{-4}$ and $\alpha_\mathrm{dz}=10^{-5}$, respectively. 

In order to compare the spatial extensions of the vortices in the GAP and DZE models, the radial (left) and azimuthal (right) density profiles taken across the vortex center are shown in Fig.\,\ref{fig:dens-profiles}. The radial and azimuthal extensions of the vortices,  $\Delta_\mathrm{rad}$ and $\Delta_\mathrm{azim}$, are determined by measuring the spatial extension of the vortex at the $1/e$th of the density maximum. For the GAP model, $\Delta_\mathrm{rad}\simeq2.7$\,au and $\Delta_\mathrm{azim}\simeq25$\,au (corresponding to $\sim244^\circ$ angular width), while for the DZE model, $\Delta_\mathrm{rad}\simeq4.2$\,au and $\Delta_\mathrm{azim}\simeq80$\,au (corresponding to $\sim95^\circ$ angular width). 

Although the vortex excitation model (GAP or DZE) could be distinguished by measuring the vortex aspect ratios, the precise determination of $\chi$ would require a very high resolution (i.e. $\sim0.01\arcsec$ beam size for our model assuming a 100\,pc source distance) to resolve the vortices radially. Moreover, in some particular cases, the vortex strengths are similar in the DZE and GAP models: $\chi\simeq10$ for the $\alpha=10^{-5}$ GAP and $\alpha_\mathrm{dz}=10^{-5}$ DZE models, respectively. Therefore, vortex aspect ratio is not an ideal parameter to use for distinguishing the two formation scenarios. However, vortices are about an order of magnitude wider in the azimuthal than in the radial direction, easing the determination of the vortex azimuthal width. The azimuthal extensions of the DZE vortices are found to be exceed $\sim180^\circ$, while they are only $\sim90^\circ$ for the GAP vortices, in agreement with the literature \citep{Lyraetal2015,Mirandaetal2016,Mirandaetal2017,ZhuBaruteau2016}.  Note that \citet{LyraMacLow2012} and \citet{Flocketal2015} presented azimuthally concentrated DZE vortices presumably caused by MHD effects, but see \cite{ZhuStone2014}. Note also that the vortex azimuthal extension in the GAP models that includes disk self-gravity is found to be in the range of $90^\circ-180^\circ$ \citep{LinPapaloizou2011}. However, the stretching requires a more massive disk than what we assumed. Therefore, we suggest that the vortex azimuthal extension is a key parameter that can be used to infer the vortex excitation mechanism as long as disk self-gravity is negligible.

Here we have to note that vertical shear instability \citep{Nelsonetal2013} is effective in transporting of angular momentum in the disk causing $\alpha\simeq10^{-4}$ \citep{StollKley2014}. Thus, the magnitude of viscosity, $\alpha=10^{-5}$, required for the long-term vortex in the GAP models is unreasonably small.  However, the $\alpha_\mathrm{dz}=10^{-4}$ is a reasonable value for the disk viscosity in the dead zone. For this reason, we compare the observational properties of vortices formed in the $\alpha=10^{-5}$ GAP and $\alpha_\mathrm{dz}=10^{-4}$ models.

\section{Synthetic ALMA Images}

The differences in the density enhancement and the spatial extent of the vortices between the DZE and GAP models may provide a way to observationally study the origin of the vortices. Submillimeter continuum observation is a straightforward choice to observationally study the vortex morphology, as most protoplanetary disks are thought to be optically thin at these wavelengths; thus, the submillimeter continuum emission probes the total surface density of the dust. \footnote{Note that the vortex center can be optically thick due to significant dust accumulation. This is the reason why we investigate not only optically thin but also thick emission.} In this section, we present synthetic observations for the Atacama Large Millimetre and Submillimeter Array (ALMA) and investigate whether or not we can infer the formation mechanism of a large-scale vortex (gap outer edge vs. dead-zone outer edge) from the morphologies seen in the synthetic observations. 

In order to study the morphology of the vortices in submillimeter observations, we assume that the mm-sized dust distribution follows that of the gas given by the hydrodynamic simulations.  However, the dust inside the planetary orbit is artificially removed in the GAP models, and complete dust clearing of mm-size dust is assumed inside the dead zone in the DZE models. In the following section, we present the plausibility of this approximation.

\subsection{Dust Depletion in the Inner Disk}
\label{sect:dust-clearing}

In the GAP model, it is assumed that the dust is completely removed because the planet clears all the material inside its gap. However, a giant planet opens only a narrow gap whose width is $\Delta R_\mathrm{GAP}\simeq5-6R_H$, where $R_H=a(q/3)^{1/3}$ is the radius of the planetary Hill sphere \citep{Cridaetal2006}. Therefore, a system of multiple planets is required to completely clear the inner disk \citep{Dodson-RobinsonSalyk2011,Zhuetal2011}. In the hydrodynamic simulations of \citet{Isellaetal2013}, three equal-mass giant planets ($q=1.9\times10^{-3}$) orbiting at $a=21,\,34$ and $55$\,au were assumed to explain the observed brightness asymmetry of LkH$\alpha$\,330. They found complete gas clearing in their simulations due to overlapping of planetary gaps. We repeated their calculations with $\alpha=2\times10^{-3}$ and $2\times10^{-4}$ (we use the same $\alpha$ assumptions for the viscosity as \citet{Isellaetal2013} are used), and confirmed the complete gas clearing. Note, however, that the vortex lifetime exceeds $10^{5}$\,yr only for the nearly inviscid case, where $\alpha$ is on the order of $10^{-5}$.

We emphasize that, according to our simulations, the observed vortex structures in the gas, i.e. the azimuthal extension and contrast, for multiple- and single-planet setups are similar.  This can be explained by the fact that RWI excitation at the outer edge of a common gap is driven  by the outermost giant planet only, while the inner planets do not affect the vortex formation there. Although our hydrodynamic models with a single giant planet (presented in Section\,\ref{sect:GAP-model}) cannot explain the inner disk clearing, assuming complete dust clearing inside the planetary orbit fairly models the formation and evolution of the large-scale vortex for multiple-planet systems.

In the DZE model, however, the inner disk is not cleared in the gas; therefore, to test the dust removal assumption, we investigated the dust drift in a viscously evolving 1D disk model. The radial drift of the solid particles in the viscously evolving dead zone is modelled in a representative particle approach, in which dust coagulation and fragmentation are not included. We followed the dust drift for $3\times10^{5}$\,yr with $10^{5}$ dust particles. In this simple 1D model, we assume that the disk is axisymmetric, thus the evolution of the surface density of the gas ($\Sigma_g(R,t)$) can be  given by
\begin{equation}
	\frac{\partial \Sigma_g(R,t)}{\partial t} = \frac{3}{R} \frac{\partial}{\partial R} \left[ R^{1/2} \frac{\partial}{\partial R} \left( \nu(R) \Sigma_g(R,t) R^{1/2} \right) \right],
	\label{eq:ido}
\end{equation}
where $\nu(R) = \alpha \delta_\alpha(R) c_s^2(R)$ is the kinematic viscosity of the gas in our dead-zone edge model, where $\delta_\alpha(R)$ is defined by equation\,(\ref{eq:deltaalpha}). 
 
 \begin{figure*}
	\centering
	\includegraphics[width=\columnwidth]{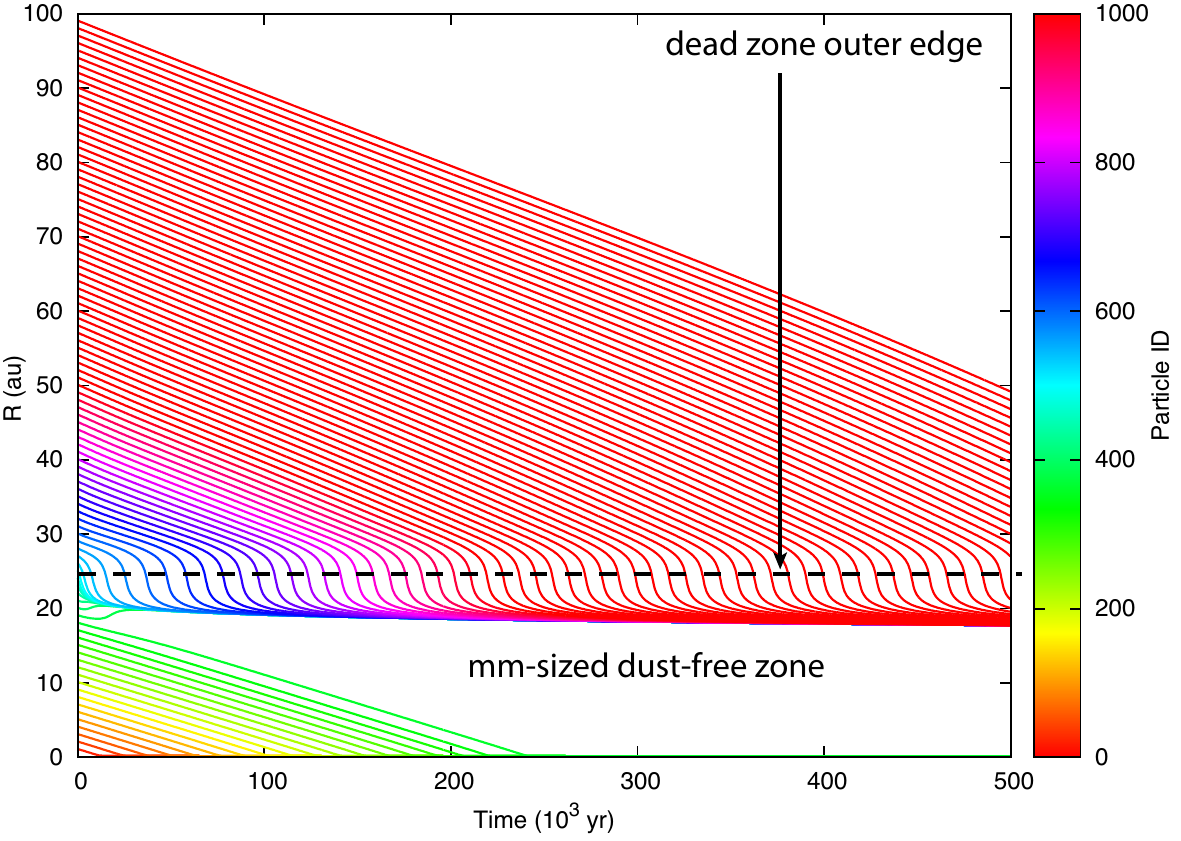}
	\includegraphics[width=\columnwidth]{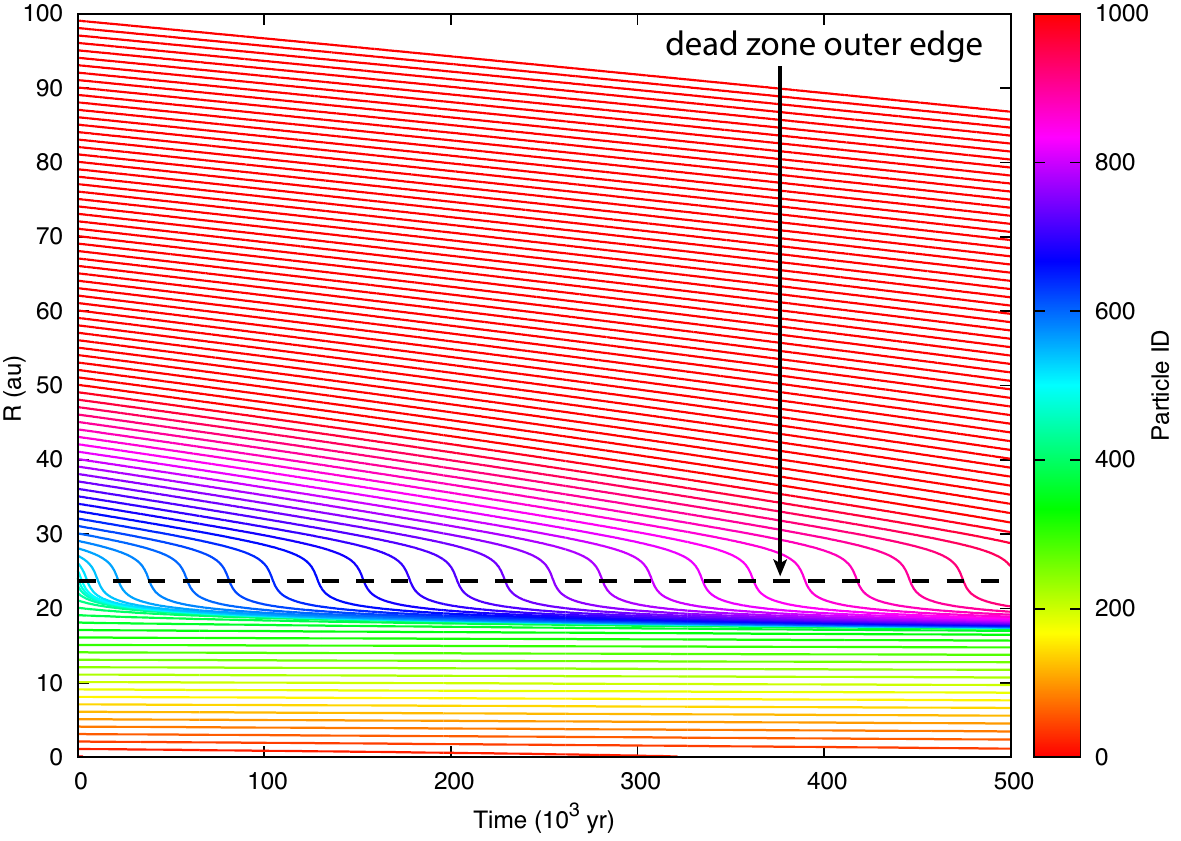}
	\caption{Radial drift of 1mm (left) and $1\mu$m (right) dust particles calculated with the 1D dust transport model assuming $\alpha=10^{-2}$, $\alpha_\mathrm{dz}=10^{-4}$, $R_\mathrm{dze}=24$\,au, $\Delta R_\mathrm{dze}=1.5H_\mathrm{dze}$, and the initial surface mass density of gas is $\Sigma_\mathrm{gas}(R)=\Sigma_0 R^{-0.5}$, where $\Sigma_\mathrm{g}=1.45316\times10^{-5} M_\odot\,\mathrm{au}^{-2}$. We show $10^2$ representative dust particles with different initial orbital distances using different colors. The dead zone becomes free of mm-size dust by $\sim250\times10^{3}$\,yr, while it remains populated with $\mu$m-sized dust.}
	\label{fig:radial-drift}
\end{figure*}

The coupling of the dust to the gas can be described by the Stokes number ($St$), which in the Epstein regime (where the interaction between particles and a single gas molecule becomes important, as the particle size is comparable to the mean free path of molecules)  is
\begin{equation}
	St = \frac{a \rho_s }{\Sigma_g(R,t)}\frac{\pi}{2},
	\label{eq:stokes}
\end{equation}
where $a$ is the size (assumed to be mm and $\mu$m in two simulations, respectively) and the intrinsic density of the solid particle is set to $\rho_s=1.6\,\mathrm{g\,cm^{-3}}$. The radial velocity of the solid particles ($u_\mathrm{dust,r}$) is given by
\begin{equation}
	u_\mathrm{dust,r} = \frac{u_\mathrm{gas,r}}{1+St^2}+\frac{2}{St+St^{-1}}u_\mathrm{drift},
	\label{eq:drift}
\end{equation}
where
\begin{equation}
	u_\mathrm{gas,r} = - \frac{3}{\Sigma_g(R,t) R^{1/2}} \frac{\partial}{\partial R} \left( \Sigma_g(R,t) \nu(R) R^{1/2} \right)
	\label{eq:vrg}
\end{equation}
is the radial velocity of the gas and
\begin{equation}
	u_\mathrm{drift} = \frac{c_s^2(R)}{2 \Omega(R)R} \frac{d\ln{P}}{d\ln{R}}
	\label{eq:drift}
\end{equation}
is the maximum drift velocity of a particle, where $\Omega(R)=R^{-3/2}$  (see, e.g.,\citealp{Weidenschilling1977,Nakagawaetal1986,YoudinLithwick2007} and \citealp{Birnstieletal2012b}). 

In our dust drift model, the the disk extends from 0.1 to 50\,au with 1500 radial cells. Similar to our fiducial 2D DZE hydrodynamic model, the outer dead-zone edge is set to $R_\mathrm{dze}=24$\,au, $\Delta R_\mathrm{dze}=1.5H_\mathrm{dze}$, and the initial surface mass density of the gas is $\Sigma_\mathrm{gas}(R)=\Sigma_0 R^{-0.5}$, where $\Sigma_\mathrm{g}=1.45316\times10^{-5} M_\odot\,\mathrm{au}^{-2}$ is used.

\begin{figure}
	\centering
	\includegraphics[width=0.9\columnwidth]{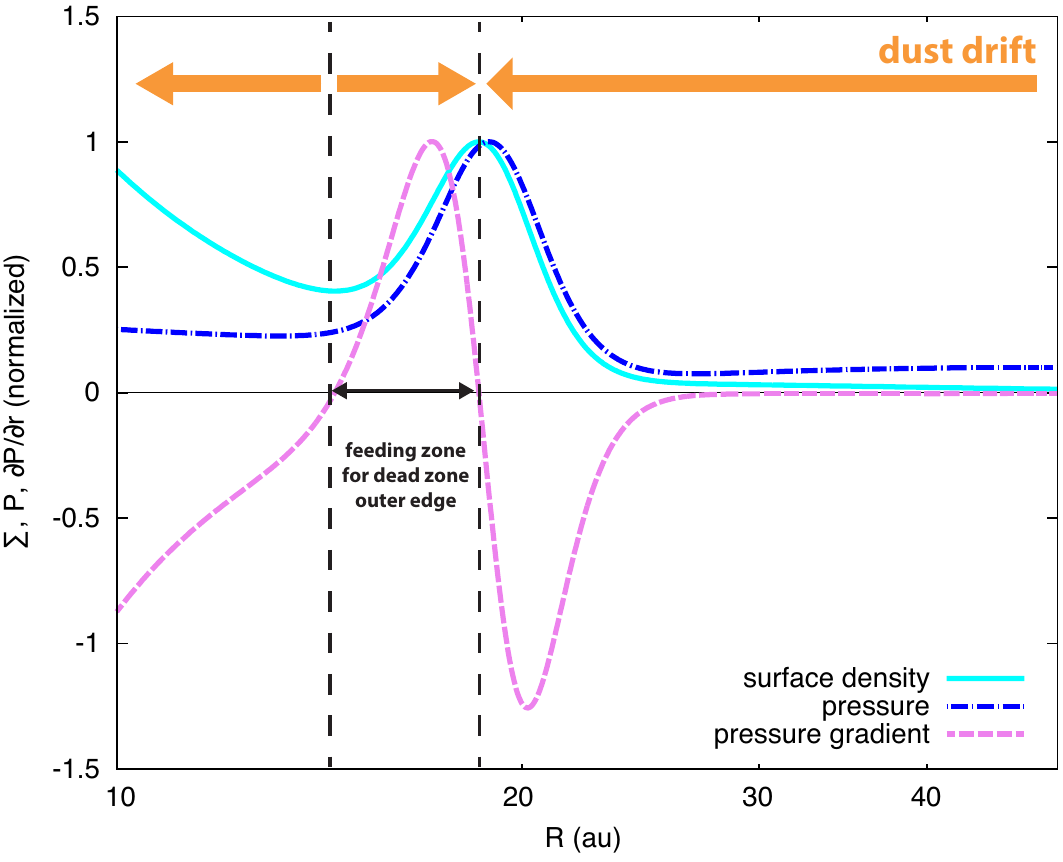}
	\caption{Azimuthally averaged density, pressure, and pressure gradient profiles in the dust drift model.  The pressure maximum impedes to refill the dead zone with dust. Inside $\sim15$\,au, the dust is subject to drift toward the central star.}
	\label{fig:dze-feeding}
\end{figure}

Fig.\,\ref{fig:radial-drift} shows the evolution of the semimajor axis of mm-sized (left panel) and $\mu$m-sized (right panel) dust particles in the fiducial DZE model. A negative pressure gradient results in sub-Keplerian gas velocity and causes rapid inward drift of dust particles, while the gas is super-Keplerian at positive pressure gradients, which causes  outward drift of dust particles (see Fig.\,\ref{fig:dze-feeding}). As a result, most of the mm-sized dust particles initially orbiting inside the dead zone drift toward the central star, while the dust initially orbiting outside the dead-zone edge are trapped at the pressure maximum. As a result, mm-sized dust particles are accumulated at the local pressure maximum. However, $\mu$m-sized dust particles are well coupled to the gas; therefore, the dead zone remains well populated with $\mu$m-sized dust.

The radial extension of the feeding zones for the pressure maxima evolves in time, which can affect their dust accumulation efficiency. Due to the viscous evolution of the disk, the pressure maximum shifts inward (see e.g. \citealp{Regalyetal2012}). We found that the mm-sized dust particles are completely absent in the dead zone by $\sim250\times10^3$\,yrs, while the dead zone is still populated by $\mu$m-sized dust. Note that if we would take into account the dust coagulation, the average dust size increases in time. Since the dust drift velocity increases with the particle size, the dust clearing would be accelerated.

Note that \citet{Birnstieletal2012a} found that the presence of the dead zone fails to explain the dust clearing of the inner disk observed in transition disks using a more elaborate model that also takes into account dust coagulation and fragmentation. However, they did not take into account the effect of the pressure maximum formed at the dead-zone outer edge. Since particles (less than several meters in size) strongly coupled to the gas cannot cross the pressure bump (Fig.\,\ref{fig:dze-feeding}), the dead zone cannot be replenished with mm-sized dust from the outer disk.

In short, it is a plausible assumption that the inner disk is completely cleared of mm-sized dust within several $10^5$\,yr for the DZE models assuming our fiducial disk parameters, while multiple planets are required to completely clear the inner disk of mm-sized dust for the GAP models.

\subsection{Radiative transfer setup for synthetic image calculation}

Synthetic images are calculated by the 3D radiative transfer code RADMC-3D\footnote{http://www.ita.uni-heidelberg.de/dullemond/software/radmc-3d}. Our imaging simulations contain three steps. First, the dust temperature is calculated in a thermal Monte Carlo simulation, then images at 880\,$\mu$m are calculated using the ray-tracing module of RADMC-3D, snd finally, ALMA observations are simulated using the Common Astronomy Software Applications (CASA) package\footnote{http://casa.nrao.edu/index.shtml} v4.2.2. 

Since FARGO solves dimensionless equations, the simulations can be scaled in terms of spatial extent and also in mass in the locally isothermal approximation. Therefore, to make our synthetic images comparable to observations, we can scale the hydrodynamic simulations such that the disk extends from 9 to 150\,au with the planetary orbital radius being at 30\,au. In this case, the vortex is located at about 30--45\,au from the star, similar to the observations (see, e.g., \citealp{Brownetal2009}), and the simulation time corresponds to $\sim1$\,Myr. For both the DZE and GAP models, we run two set of simulations, where the core of the vortex is optically thin and thick (by appropriately scaling the density).  

In the RADMC-3D simulations, we use a 3D spherical mesh ($R, \phi, \theta$). The radial grid extends between 9 and 150\,au, and contains $N_R=256$ grid cells. Along the azimuthal and poloidal angular coordinates, we use $N_\phi=512$ and $N_\theta=180$ grid cells, respectively. While the azimuthal grid is equidistant, the grid cells in the poloidal coordinate are distributed such that we use 10, 160, and 10 grid points in the [0, $\pi/2-0.25$], [$\pi/2-0.25$, $\pi/2+0.25$], and [$\pi/2+0.25$, $\pi$] intervals, respectively. In each poloidal interval, we use a uniform, equidistant grid. 
 
Hydrodynamic frames are interpolated to the RADMC-3D mesh and converted to volume density assuming a Gaussian vertical density distribution assuming a constant aspect ratio of $h=0.05$, similar to the 2D hydrodynamic simulations. In the radiative transfer model, we use the optical constants of the astronomical silicate from \citet{WeingartnerDraine2001} for the dust particles and calculate the absorption and scattering cross-sections using Mie theory. Dust grains in the model have a size distribution of $n(a)\propto a^{-3.5}$ between $0.1$ and 1000\,$\mu$m, corresponding to the steady-state mass distribution of the collisional fragmentation cascade \citep{Dohnanyi1969}. The central radiation source in the model has parameters representative of that of a Herbig\,Ae star: R$_\star$=2.5\,$R_\odot$, T$_\star$=9500\,K, M$_\star$=2.0\,$M_\odot$.

To determine the dust temperature of individual dust grains, we run a thermal Monte Carlo simulation using 1.2$\times10^8$ photon packets. The continuum images at 880\,$\mu$m are calculated with ray tracing, consisting of 800 by 800 pixels, with a pixel size of 0.89\,mas assuming a source distance of 140\,pc. 

Simulated ALMA observations are calculated with the CASA {\tt simobserve} task to generate synthetic visibilities and the {\tt clean} task for imaging. The full 12m array was used for the simulated observations, in three different configurations resulting in a spatial resolution of approximately 0\arcsec.25", 0\arcsec.18, and 0\arcsec.1. We assume the full 7.5\,GHz continuum bandwidth of the ALMA correlator and calculate the visibilities for a total integration time of 30 minutes. The simulations include thermal noise arising from a water vapor column of 0.913\,mm representative of Band\,7 observations. 

In both optically thin and thick cases, we calculated images at two inclination angles, $30^\circ$ and $60^\circ$ and five different azimuth angles for the vortex between 0$^\circ$ and 180$^\circ$ in 45$^\circ$ steps. To study the effect of the shape of the synthesized beam on the morphology of the images, we simulated observations assuming a source decl. of 25$^\circ$ and 70$^\circ$, and the times of the observations are centered on 0, -1, and -2\,hr angles.

\subsection{Analysis of Synthetic Images}

\begin{figure*}
	\centering
	\includegraphics[width=19cm]{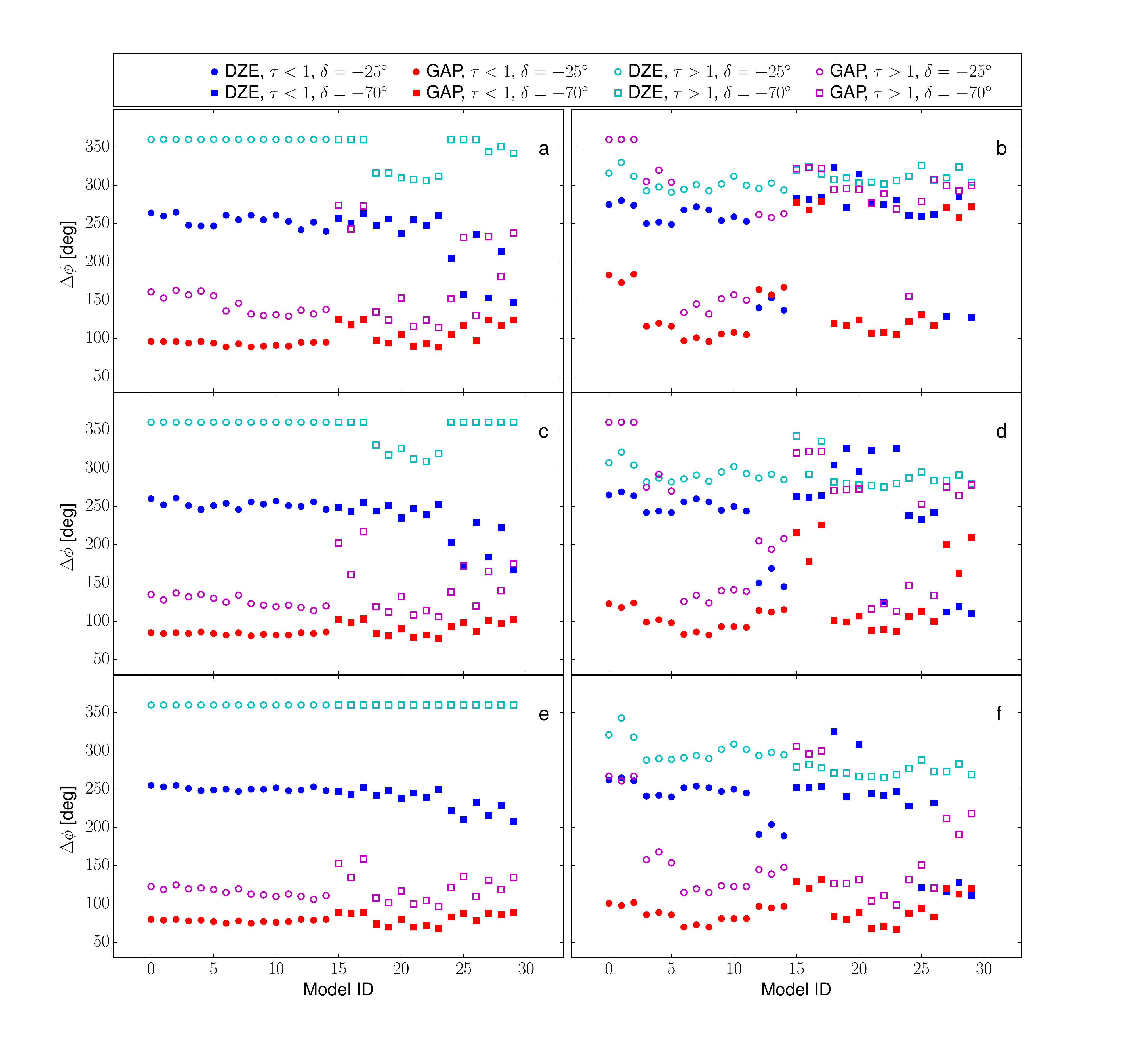}
	\caption{Azimuthal extent of the vortices as measured in the synthetic ALMA images at spatial resolutions of 0\arcsec.25 (panel (a, (b)), 0\arcsec.18 (panel (c), (d)), 0\arcsec.1 (panels (e), (f)). The left column shows models for an inclination angle of 30$^\circ$, while the right column shows the models for 60$^\circ$. The filled and open symbols represent optically thick and thin models, respectively. The Model ID represents the combination of five different position angles of the vortex, three different hour angles (0, -1, and -2\,hr) the observations are centered on, and two different optical depths at the core of the vortex (optically thin and thick).}
	\label{fig:vortex_size}
\end{figure*}

The most striking difference between the vortices excited at the edge of a planet-carved gap and at the edge of a dead zone is  their azimuthal extent (see Sect.\,\ref{sect:hydro-morph}). Our hydrodynamic simulations show that vortices formed at the edge of a planetary gap are significantly more compact in the azimuthal direction than their counterparts at the edge of a dead zone. In the following, we show that this property can be used in observations to infer the formation mechanism of the vortices.

To test the applicability of this technique, we determine the azimuthal extent of the vortices in the synthetic ALMA observations in the following way. As a first step, we deproject the images with a known inclination and position angle. In real observations of well-studied objects, both the inclination and the position angle are known to a reasonable (several degrees) accuracy from gas-line observations. Then, we converted the image from Cartesian to polar coordinates centered on the stellar position. Finally, we measured the azimuthal size of the region where the intensity is 66\% of that of the peak intensity in the core of the vortex. The measured sizes of the vortices are presented in Fig.\,\ref{fig:vortex_size} for different position angles of the vortex, hour angles the observations are centered on, and optically thick and thin vortex core assumptions.

For low inclination angles ($30^\circ$; panels (a), (c), and (e) of Fig.\,\ref{fig:vortex_size}), the DZE and GAP models can be separated in most cases based on the measured azimuthal size of the vortex. While for the lowest resolution ($\sim$0\arcsec.25; see panel (a)), some confusion may occur between the two models, at the highest resolution (0\arcsec.1; see panel (e)) the DZE and GAP models are clearly separated. For high inclination angles ($60^\circ$; panels (b), (d), and (f) of Fig.\,\ref{fig:vortex_size}), separating the two models becomes more challenging, especially for the lowest spatial resolution. 

Based on our results presented in Fig.\,\ref{fig:vortex_size}, we conclude that if (1) the disk is optically thin, (2) the inclination is low (probably 45$^\circ$ or less) and (3) the resolution is sufficiently high (the beam size is at least twice as small as the projected distance from the star to the vortex core), the DZE and GAP models can be distinguished. As we show in the following, the morphology of the vortex in the two different models  can be very similar for higher optical depths and inclination angles and lower spatial resolutions.

\subsubsection{Effect of optical depth}

\begin{figure}
	\centering
	\includegraphics[width=9cm]{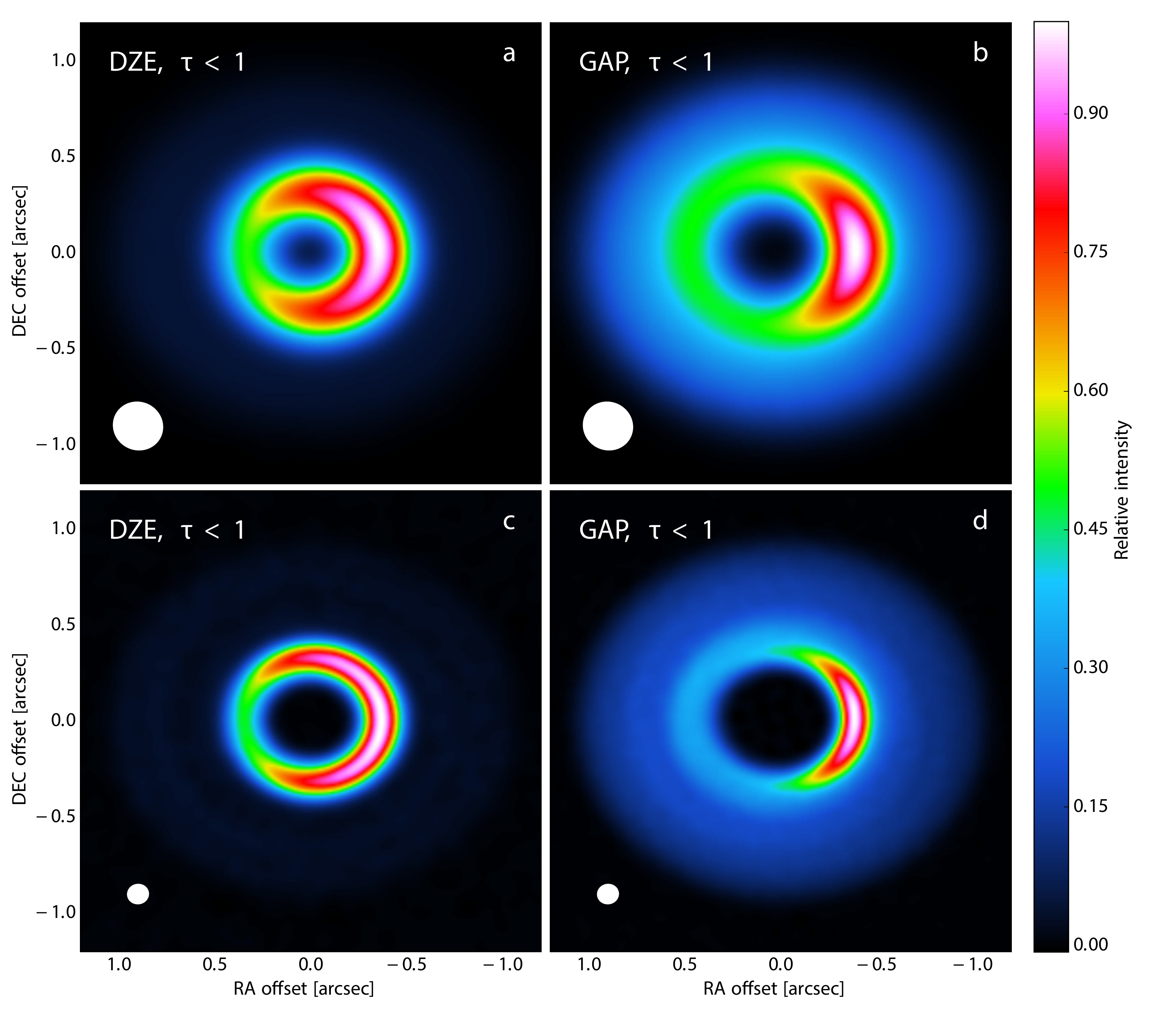}
	\caption{Synthetic ALMA observations of optically thin DZE (panels (a), (c)) and GAP (panels (b), (d)) models at two different resolutions (0\arcsec.25 on panels (a), (b) and, 0\arcsec.1 in panels (c), (d)). The arc of the vortex is significantly more extended in the azimuthal direction for a DZE model compare to a GAP model.}
	\label{fig:clean_alma_prediction_i30}
\end{figure}

\begin{figure}[!h]
	\centering
	\includegraphics[width=9cm]{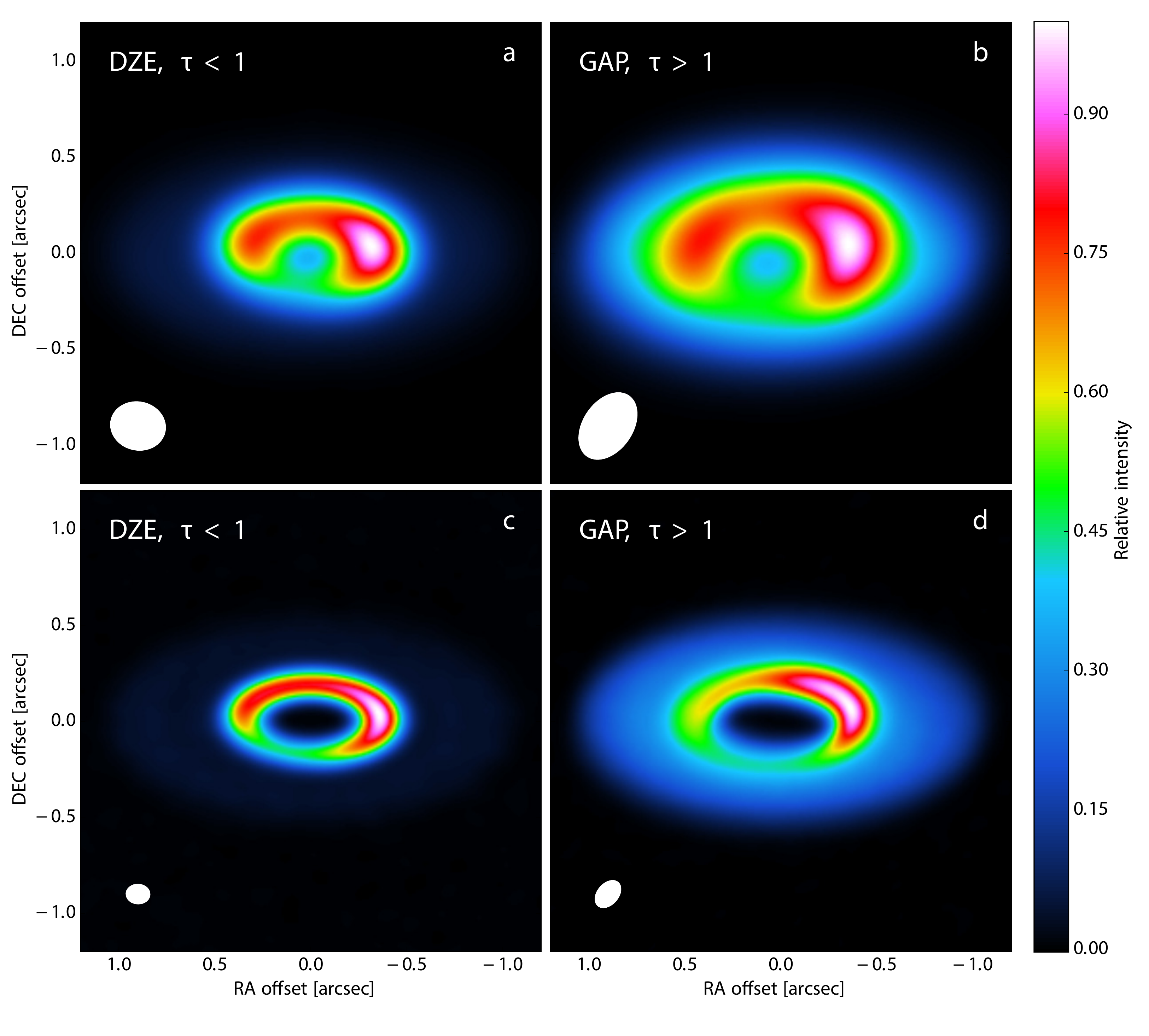}
	\caption{Synthetic ALMA observations of an optically thin DZE (panels (a), (c)) and an and optically thick GAP (panels (b), (d)) models at two different resolutions (0\arcsec.25 in panels (a), (b) and 0\arcsec.1  in panels (c), (d)). Once the disk becomes optically thick at the edge of the gap, the arc of the vortex also appears azimuthally extended in a GAP model; however, the models can be distinguished at high resolution.}
	\label{fig:optical_depth_effect_i60}
\end{figure}

The true azimuthal extent of the vortex can only be measured in submillimeter continuum images if the disk is optically thin, in which case the DZE and GAP models can easily be distinguished (Fig.\,\ref{fig:clean_alma_prediction_i30}). Once the disk becomes optically thick, the surface brightness is  determined by the temperature instead of the surface density. Since the highest surface density is in the core of the vortex, the disk optical thickness becomes the highest there. The size of the region where the observed intensity drops below a certain fraction of the peak intensity (i.e. the azimuthal extension of the horseshoe-shaped asymmetry) also increases with the optical depth. Therefore, the azimuthal extension of the vortex is overestimated if the disk is optically thick at the vortex core. For the optically thick case, the azimuthal contrast across the vortex is also decreased by the saturation of the emission at the vortex core. The combination of these effects can make the appearance of a vortex at a planetary gap edge very similar to that of a vortex at the edge of a dead zone, but only for low resolution (Fig.\,\ref{fig:optical_depth_effect_i60}).

\subsubsection{Effect of beam shape}

\begin{figure}[!h]
	\centering
	\includegraphics[width=9cm]{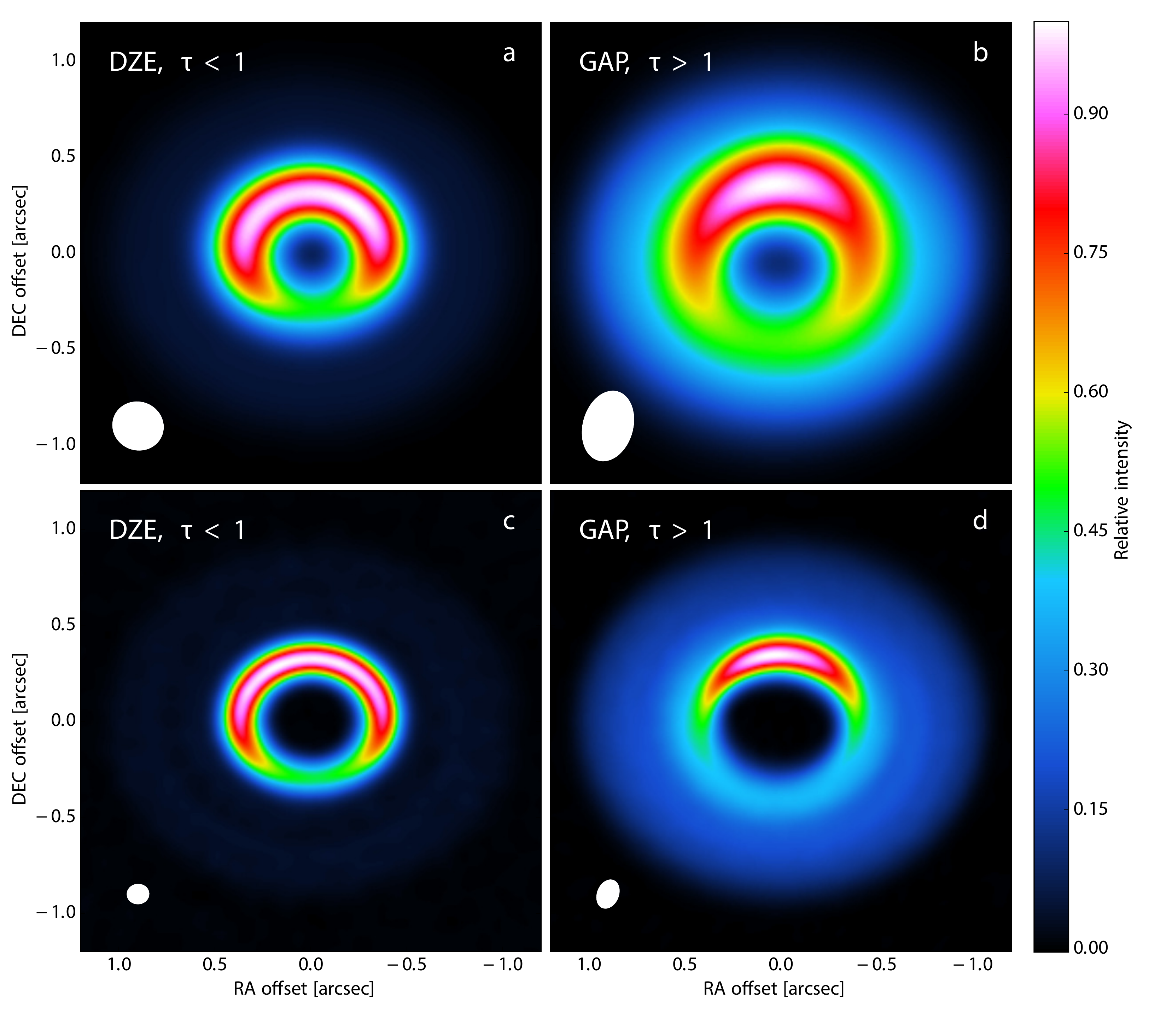}
	\caption{Same as Fig.\,\ref{fig:optical_depth_effect_i60} but for an inclination angle of 30$^\circ$. If the synthesized beam is elongated and the major axis of the beam ellipse aligns with the position of the vortex, the azimuthal extent of the vortex in the GAP model (panel (b)) can look similar to that of a vortex in the DZE model  (panel (a)). Increasing the spatial resolution can break the degeneracy in the apparent vortex morphology, and the two models can be distinguished (panels  (c), (d)).}
	\label{fig:optical_depth_effect_i30}
\end{figure}

In interferometric observations, the synthesized beam is not generally circular, but rather elliptic due to the combination of the elevation of the target
above the horizon and the antenna configuration. The ellipticity of the beam causes the structures in the image to look more elongated along the direction of the beam major axis. Such elliptical beams can cause a characteristic changes to the apparent morphology of the observed vortices.

The azimuthal extent of the vortex in the GAP models depends on the position angle of the beam ellipse. The apparent azimuthal extent of the vortex is the highest if the beam major axis is perpendicular to the vortex. For the optically thick case, the vortex in the GAP model can look azimuthally as much extended as a DZE vortex (panels (a) and (b) of Fig.\,\ref{fig:optical_depth_effect_i30}). 

At low inclination angles, the peak of the emission is located in the middle of the extended horseshoe-shaped structure, i.e. at the vortex center. At high inclination angles, however, the peak of the brightness can shift from the vortex center, depending on the relative position of the beam major axis and the vortex (panels (b) and (d) in Fig.\,\ref{fig:optical_depth_effect_i60}). We note that the horseshoe-shaped asymmetry is always azimuthally less extended in the GAP models than in the DZE models in the optically thin case (Fig.\,\ref{fig:optical_depth_effect_i60}).

\begin{figure}
	\centering
	\includegraphics[width=9cm]{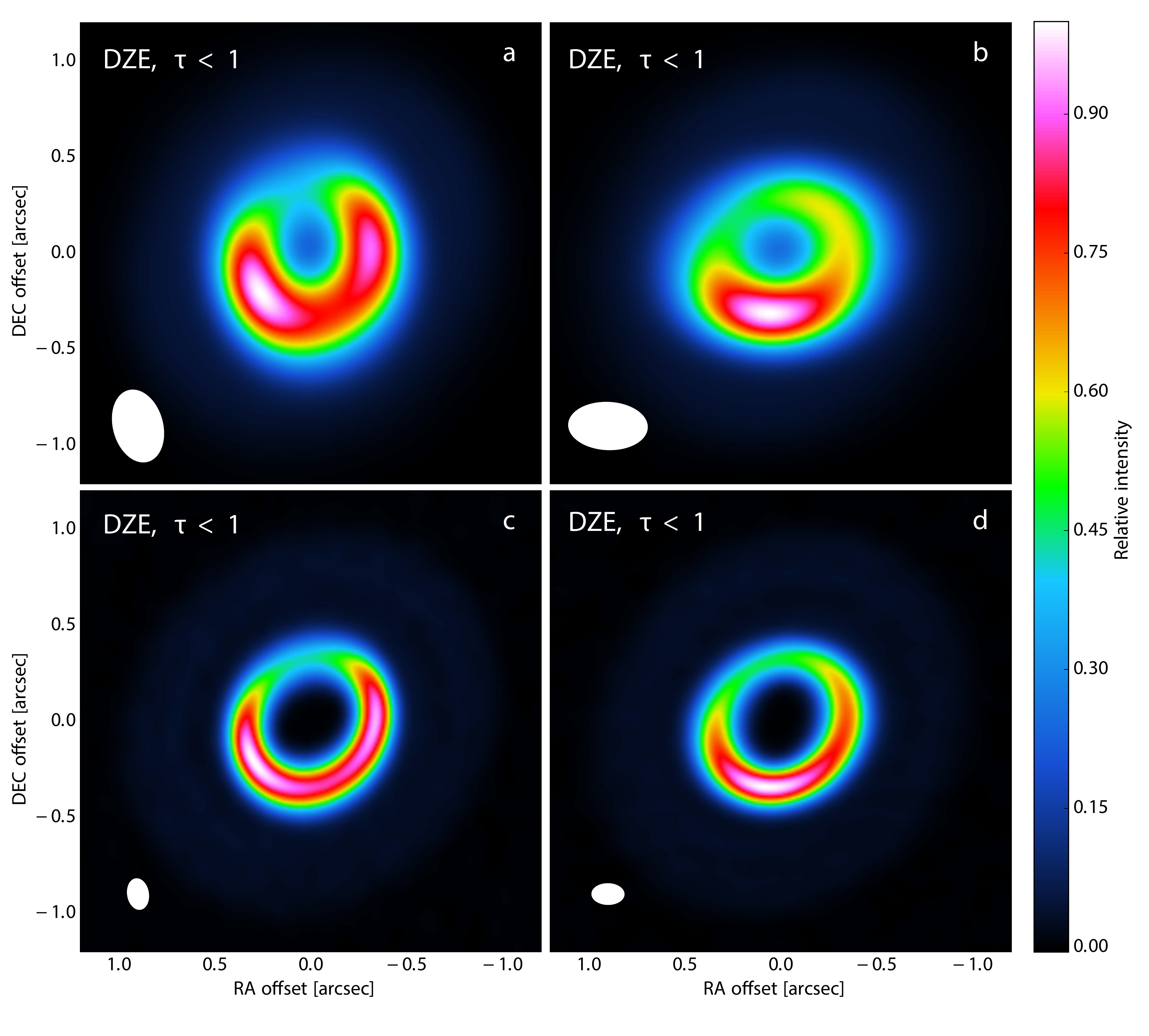}
	\caption{Effect of beam shape on the image of a DZE vortex. The same radiative transfer image was used to generate the synthetic observations. To change the shape of the beam, we changed decl. of the source in the simulated observations from -20$^\circ$ (panels (a), (c)) to -70$^\circ$ (panels (b), (d)). The inclination of the disk and the position angle of the vortex were chosen to match those of SR\,21 \citep{Perezetal2014}. For elongated beams at low spatial resolution, a DZE vortex can show one or two distinct peaks, depending on whether the major axis of the beam aligns with the position of the vortex (panel (a)) or perpendicular to it (panel (b)). At high spatial resolution, this effect is also present (panels (c), (d)). }
	\label{fig:beamsize_effect}
\end{figure}

In contrast to the GAP models, the vortex in the DZE models can split into two bright blobs if the beam is elongated. For optically thick DZE vortices, this is always the case. If, however, a DZE vortex is optically thin, the number of bright blobs and/or the azimuthal extent of the observed horseshoe depends strongly on the position angle of the beam ellipse with respect to that of the vortex. If the beam major axis is perpendicular to the vortex, an azimuthally extended horseshoe appears with two distinct bright peaks (panel (a) of Fig.\,\ref{fig:beamsize_effect}). In contrast, if the beam major axis is parallel to the vortex, a single-peaked  asymmetry appears with a significantly smaller azimuthal extension (panel (b) of Fig.\,\ref{fig:beamsize_effect}). 

We note that this direction-dependent "smearing" effect also depends on the ratio of the structure size to the beam size. Beam-shape effects cannot change the morphology of structures in the image that are larger than a few beam sizes. Therefore, increasing the angular resolution in the observations can in most cases break the degeneracy, making it possible to distinguish DZE and GAP models based on the vortex azimuthal extent.

\section{Discussion}

\subsection{Comparison to observations}

While there is now observational evidence for nonaxisymmetric surface brightness distribution in the submillimeter for more than a dozen transitional disks, the precise morphology of the asymmetry is rather uncertain in many cases due to insufficient spatial resolution and/or the low signal-to-noise ratio of the observation. However, in a few cases, if ALMA observations are available and the structure of the disk is known well enough, that we can qualitatively compare the observed morphology of the disk --- most importantly, the azimuthal extent of the observed vortices --- to our models. 

There are four well-known transitional disks with ALMA observations and evidence for a vortex in the disk: Oph\,IRS\,48  \citep{vanderMareletal2013}, HD\,142527 \citep{Casassusetal2013}, HD\,135344B and \citep{Perezetal2014}, and SR\,21 \citep{Brownetal2009}. \citet{vanderMareletal2013} explained the extreme strong asymmetry in the disk of IRS\,48 with a vortex excited by a massive ( $>10\,M_{\rm Jup}$) giant planet. \citet{vanderMareletal2013} et al. presented a numerical model of a gap-edge vortex showing remarkable similarities with the observations. 

The azimuthal extent of the asymmetries in the other three sources, HD\,142527, HD\,135344B, and SR\,21, seems to be significantly larger than that in the disk of Oph\,IRS\,48 and thus may not necessarily be consistent with a planet-induced vortex. To fit the observed images, \citet{Perezetal2014} assumed a superposition of a planet-induced vortex and a full ring for both HD\,135344B and SR\,21. For canonical disk viscosity values ($10^{-3}\leq\alpha\leq10^{-2}$), only a symmetric dust ring is expected to form at a planet-carved gap edge. Therefore, such azimuthally extended asymmetries can be better explained by a sudden jump in the viscosity (proposed by \citet{VarniereTagger2006} and later compared to observations by \citet{Regalyetal2012}, rather than by putative giant planet(s).

Our models of a DZE vortex bear especially remarkable similarities with the images of SR\,21. In \citet{Brownetal2008} the image of SR\,21, obtained with the Sub-Millimeter Array (SMA), shows a horseshoe-shaped structure with more than $270^\circ$ in azimuthal extent and with two bright blobs on the two opposite sides of the arc. In contrast, the ALMA image in \citet{Perezetal2014} shows only a single maximum along the horseshoe-shaped asymmetry at the position between the two blobs seen in the SMA image. In Fig.\,\ref{fig:beamsize_effect}, we show that an azimuthally extended DZE vortex can reproduce the morphology of both the SMA and the ALMA images remarkably well if we take into account the differences in the shapes of the synthesized beams in the two observations. Note that since our goal was to show a qualitative comparison of our models with the observations, we did not use the exact same uv-coverage as in the real SMA and ALMA observations of SR\,21. Instead, we changed the decl. of our target in the synthetic observations such that the resulting synthesized beam will have a similar shape to that in the real observations. 

So far, horseshoe-shaped asymmetries in protoplanetary disks have been interpreted with vortices induced by planets. Here we suggest that in a few cases where the vortex is azimuthally extended, this may not be the case, and the vortex could be induced by a viscosity jump \citep{Regalyetal2012}. Unfortunately, the images obtained so far do not have high enough spatial resolution that an azimuthally extended vortex could be distinguished from other models, e.g., a planet-induced vortex + ring proposed by \citet{Perezetal2014}. A significantly higher spatial resolution (by a factor of about 2-3 at least) compared to already existing one is needed to unambiguously identify the presence of such extended vortices.

\subsection{Outlook and Caveats}

Here we have to mention some caveats of our model. Regarding the hydrodynamical simulations, the most obvious caveat is that we run simulations in 2D. However, \citet{Meheutetal2012c} found that the vortex formation and development in 3D are indeed very similar to that of 2D: the disk vertical stratification only slightly decreases the growth rate of vortices compared to that of 2D. 

We neglect the effect of disk self-gravity, while \citet{LinPapaloizou2011} found that in self-gravitating disks, the gap-edge vortex coagulation ends at mode $m=2$ for relatively high disk masses ($\geq0.031M_*$). As a result, two vortices may present simultaneously in the disk. Thus, in the GAP models, we should see a double-peaked submillimeter brightness distribution similar to that of in the DZE models. \citet{LovelaceHohlfeld2013} found that the disk self-gravity could be important regarding the vortex formation if $Q<R/h$. Recently, \citet{ZhuBaruteau2016} found that taking into account the disk's self-gravity results in weakened vortex, especially for massive disks. Note, however, that they artificially imposed a density jump on the disk to form a large-scale vortex. Therefore, DZE or GAP vortex formation in self-gravitating disks assuming modest disk mass is  worth investigating (see, e.g. \citealp{RegalyVorobyov2017b}).

We neglect the planet growth in the GAP models. Recently, \citet{Hammeretal2017} found that the vortex azimuthal extension generated by a slowly growing giant planet ($T_\mathrm{growth}$ being on the order of several 1000 planetary orbits) can be larger by about a factor of two, similar to that of the DZE models. Note, however, that they prescribed planet growth by an artificial sinusoidal function (see their Equation~(1)) neglecting the viscous dynamics inside the planet Hill sphere. Thus, it is worth modeling RWI excitation with a more elaborate accretion prescription,  like what is described in \citet{Kley1999}.

We assume a locally isothermal disk, which is valid only if the cooling time of the gas is fast compared to the local orbital time-scale. This assumption, however, can be plausible for a several-million-years-old transition disk. Nevertheless, it is worth investigating vortex formation and survival in adiabatic disks, as \citet{Lyraetal2009b} found stronger RWI excitation in adiabatic disks, which may have consequences for the vortex evolution.

We completely neglect the MHD effects in our simulations. However, \citet{ZhuStone2014} showed that 2D simulations assuming a kinematic viscosity (approximated to the stresses of their 3D MHD models) show vortex formation that is very similar to their full MHD simulations. Note also that a long vortex lifetime in the 3D MHD simulations of \citet{ZhuStone2014} was only observed if a nonideal effect, such as the ambipolar diffusion, was taken into account. The nonideal effect of ambipolar diffusion results in suppression of MRI turbulence, i.e. low disk viscosity, which helps to maintain the vortex.

In our dead-zone models, we assume quite sharp viscosity transitions to excite RWI. The simulations of \citet{Lyraetal2009b} and \citet{Regalyetal2012} agree in that RWI excitation occurs only for sharp dead-zone edges in which the width of the viscosity transition is smaller than twice the local pressure scale height. However, \citet{Dzyurkevichetal2013} have shown that resistivity changes very smoothly (therefore, MRI generated viscosity parameter $\alpha$ too) in the outer dead-zone edge. As a result, the dead-zone edge in an $\alpha$ disk is about five times thicker than the required threshold value for RWI excitation. However, replacing the $\alpha$ prescription with resistive MHD, a smooth change in gas resistivity, is found not to  imply a smooth transition in turbulent stress \citep{Lyraetal2015}. As a result, a large-scale vortex is stable even in smooth resistivity transition, and the abovementioned requirements for exciting RWI is only an inherent feature of $\alpha$ models.

Regarding the calculation of synthetic images, there are also some caveats in our model. The simple assumption of that the dust density distribution follows that of the gas (i.e. the dust-to-gas ratio is constant throughout the disk) neglects the earlier results of \citet{Johansenetal2006}, who observed strong enhancement of the dust-to-gas mass ratio in the presence of anticyclonic vortices. \citet{Birnstieletal2013} also found that the dust-to-gas mass ratio could become higher than the global value if the viscosity parameter was larger than the particle's Stokes number. Moreover, we neglect the dust feedback to the vortex evolution too. Recently, \citep{Fuetal2014b} investigated the dust feedback on the vortex excited by a giant planet having a $q=5\times10^{-3}$ planet-to-star mass ratio and found that the dust feedback can destroy the vortex if a large amount of dust ($10\,M_\oplus$) is accumulated inside the vortex. Note that \citet{Regalyetal2013} investigated the migration trapping of low-mass planets ($10\,M_\oplus$) in the presence of dead-zone edge vortex and found similar vortex destruction as the planet enters to the vortex. However, since the trapping of the low-mass planet is found to be only temporary, the vortex redevelops due to an existing sharp density jump formed at the dead-zone edge. Therefore, we think that the gap-edge vortex is vulnerable to and the dead-zone edge vortex is stable against dust feedback. In any case, to give a more reliable model for the evolution of gap-edge and dead-zone-edge vortices, both the dust and gas dynamics should be incorporated in hydrodynamic models.

\section{Summary and Conclusion}

The conventional explanation for the large-scale brightness asymmetries of transition disks is the accumulation of dust grains in the core of anticyclonic vortices. Such vortices can form by the excitation of RWI in the vicinity of steep pressure gradients formed, e.g. at the edges of a giant planet-carved gap (GAP model) or the outer edge of a disk's dead zone (DZE model). We investigated by means of 2D hydrodynamical simulations whether or not we can infer the formation mechanism of the vortex from the morphology of the disk as seen in submillimeter ALMA observations.

We studied GAP and DZE vortex formation mechanisms by means of 2D locally isothermal hydrodynamical simulations with an $\alpha$-type prescription for the disk's viscosity. We modeled the vortex formation at the outer edge of an embedded giant planet-carved gap ($R_p=10$\,au, $q=1.25,\,2.5,\,5,$ and $10\,M_\mathrm{Jup}$) in three different viscosity regimes ($\alpha=10^{-3},\, 10^{-4},\, 10^{-5}$) and in a disk that has a sharp dead-zone edge ($\alpha=10^{-2}$, $\alpha_\mathrm{dz}=10^{-4}$, $R_\mathrm{dze}=24$\,au, $\Delta R_\mathrm{dze}=1H_\mathrm{dze}$ and $1.5H_\mathrm{dze}$). We tested the vortex lifetime against the steepness of the initial density profiles ($\Sigma\sim R^{-p}$, where $p=0.5,\,1$, and $1.5$) and disk geometry that is flat and flared. 

Although RWI is excited and a large-scale vortex develops in both models, its survival for a sufficiently long time to observe requires nearly inviscid GAP models, $\alpha\leq10^{-5}$, while an order of magnitude lower disk viscosity, $\alpha_\mathrm{dz}\leq10^{-4}$, is sufficient for DZE models. We emphasize that 3D hydrodynamical models showed that $\alpha\gtrsim10^{-4}$ in protoplanetary disks due to the vertical shear instability \citep{StollKley2014}. Long-lasting vortex development favors a relatively smooth initial density slope, i.e. $p\leq1$ in the GAP models. The GAP and DZE vortices have distinct geometric features, which is suitable to differentiate the two formation scenarios. Our main findings based on synthetic ALMA images calculated from the hydrodynamic simulations are as follows.

(1) We showed that the GAP vortices are generally stronger than the DZE vortices; the aspect ratios are $\sim8/10$ and $\sim10/14$ for $\alpha=10^{-4}/10^{-5}$, respectively. Vortices formed in the DZE models are spatially more extended ($\gtrsim180^\circ$) than in the GAP models ($\sim90^\circ$), which can be used to distinguish the two formation scenarios.

(2) Following the dust drift in a 1D model, the disk inside the dead zone is found to be cleared of mm-sized dust within $\sim2.5\times10^5$\,yr, due to the efficient dust collection of pressure maxima formed at the dead-zone edge. Contrarily, the inner disk is still populated with $\mu$m dust.

(3) In the submillimeter, the brightness asymmetries are significantly different for the GAP and DZE models assuming optically thin disk emission. The brightness asymmetry is azimuthally more concentrated for the GAP than the DZE models: the azimuthal brightness contrast is $\sim4$ in GAP and $\sim2$ in DZE models. In the DZE models, the brightness distribution shows multiple peaks for low disk inclination angles ($i\simeq60^\circ$), due to the relatively lower azimuthal brightness contrast and projection effect.

In summary, we found a tentative evidence that the shape of the surface brightness asymmetries in submillimeter wavelengths correlates with the vortex formation process, within the limitations of the $\alpha$ viscosity approximation. One needs, however, to resolve the azimuthal extent of the vortex in the GAP models (i.e. by 4--5 resolution elements/beams). The azimuthal extent of the gap-edge vortex  is found to be approximately 2 rad (see Sec\,\ref{sect:GAP-model}). If the vortex is located at 40\,au from the star, and the star is at a distance of 140\,pc, this translates to a spatial resolution requirement of about 0\arcsec.1--0\arcsec.15.

Assuming that the $\alpha$-viscosity prescription of \citet{ShakuraSunyaev1973} is adequate for transition disks, our analysis of synthetic images (Figs.\,\ref{fig:clean_alma_prediction_i30}-\ref{fig:optical_depth_effect_i30}) suggests that the dead-zone-edge vortex scenario might be more plausible than the gap-edge vortex scenario for sources that show double-peaked or horseshoe-like brightness asymmetries (e.g. HD\,142527, HD\,135344B, and SR\,21). However, the existence of an RW unstable viscosity transition in protoplanetary disks has not yet been confirmed. For sources that show single-peaked strong brightness asymmetries (e.g. Oph\,IRS\,48), the gap-edge scenario could give a better explanation, although for conventional viscosity values ($\alpha\simeq10^{-3}-10^{-2}$), the shortlifetime of the gap-edge vortex challenges our current understanding.

\section*{Acknowledgments}

This project was supported by the Hungarian National Research, Development and Innovation Office grant No. 119993. ZsR acknowledges the support of the Momentum grant of the MTA CSFK Lend\"ulet Disk Research Group. AJ acknowledges the support of the DISCSIM project, grant agreement 341137, funded by the European Research Council under ERC-2013-ADG. ZsR gratefully acknowledges the support of NVIDIA Corporation with the donation of Tesla GPUs and NIIF for access to computational resource based in Hungary at Debrecen. A report by an anonymous referee further improved the quality of the manuscript.

%

\end{document}